\documentclass[useAMS,usenatbib,referee]{mn2e}
\usepackage{graphicx,graphics, amsmath}
\title[1A~1118-61 with $Suzaku$]{Timing and broad band spectroscopy of 1A~1118-61 with Suzaku }
\author[C. Maitra, B. Paul and S. Naik]{Chandreyee Maitra$^{1,2}$\thanks{E-mail: cmaitra@rri.res.in;} Biswajit 
Paul$^{1}$\thanks{E-mail: bpaul@rri.res.in;} and Sachindra Naik$^{3}$\thanks{E-mail:snaik@prl.res.in}\\
$^{1}$Raman Research Institute, Sadashivnagar, Bangalore-560080, India\\
$^{2}$Joint Astronomy Programme, Indian Institute of Science, Bangalore-560012, India\\
$^{3}$Astronomy and Astrophysics Division, Physical Research Laboratory, Navrangapura, Ahmedabad-380009, Gujarat, India}
\begin{document}

\date{}

\maketitle

\label{firstpage}

\begin{abstract}
 We present a timing and broad-band pulse phase resolved spectral analysis 
of the transient Be X-ray binary pulsar 1A~1118-61 observed during its outburst in January 2009 using 
\emph{Suzaku} observations. \emph{Suzaku} observations were made twice, once at 
the peak of the outburst, and once 13 days later at its declining phase. 
Pulse profiles from both observations exhibit strong energy dependence with several peaks at low 
energies and a single peak above $\sim$10 \rmfamily{keV}. A weak, narrow peak is detected at the 
main dip of the pulse profiles from both observations in the energy bands below 3 \rmfamily{keV}, 
indicating the presence of a phase dependent soft excess in the source continuum. 
The broad-band energy spectrum of the pulsar could be fitted well with a partial covering 
cutoff power-law model and a narrow iron fluorescence line. We also detect
a broad cyclotron feature at $\sim$50 \rmfamily{keV} from both observations which is a 
feature common for accretion powered pulsars with high magnetic field strengths. Pulse 
phase-resolved spectral analysis shows an increase in the absorption column density of the partial 
covering component, as well as variation in the covering fraction at the dips of the pulse profiles, that 
naturally explains the energy dependence of the same. The cyclotron line parameters also show significant variation with 
pulse phase with a $\sim$ 10 \rmfamily{keV} variation in the cyclotron line energy and a variation in depth by a factor of three. 
This can be explained either as the effect of different viewing angles of the dipole field at different pulse phases, or 
a more complex underlying magnetic field geometry.
\end{abstract}
\begin{keywords}
X-rays: binaries-- X-rays: individual: 1A 1118-61-- stars: pulsars: general
\end{keywords}
\section{Introduction}
Be/ X-ray binaries are a subtype of neutron star high mass X-ray binaries (HMXBs) where 
the optical counterpart is a dwarf, subgiant or a giant O/Be star (luminosity 
class III-V) \citet{reig2011}. These systems typically have a large orbital period, moderate 
eccentricity, and being a non-supergiant system, the Be star lies deep within the Roche 
lobe. X-ray emission is believed to be due to the accretion from the circumstellar 
disc of the rapidly rotating Be star onto the compact star. Most Be/ X-ray binaries are transient systems 
and display two types of outbursts such as (i) short, periodic, less luminous 
Type I outbursts ($ L_{x} \leq 10^{36} - 10^{37} \mathrm{erg\quad s^{-1}}$) lasting for 
a few days occurring at the phase of the periastron passage, and (ii) giant, longer, Type II outbursts ($ L_{x} \geq 10^{37} \mathrm{erg\quad s^{-1}}$) 
lasting for several weeks to months occurring when a large fraction of matter is accreted from the Be star's disk.
 Type II outbursts are sometimes followed by smaller recurrent type I outbursts.

Almost all Be/ X-ray binaries are accretion powered pulsars having a wide range of periods
from seconds to hundreds of seconds. These are highly magnetized neutron stars accreting 
matter, channeling it along magnetic fields onto the magnetic poles, and forming an accretion 
column of X-ray emitting plasma on the neutron star surface \citep {pringle1972, davidson1973, 
gnedin1973, lamb1973}. The spectrum of these objects can be described by 
Comptonization of the soft X-rays by scattering in the hot plasma \citep {barnard1981,brainerd1991,meszaros1985a,
 meszaros1985b, nagel1981a, nagel1981b}. In the presence of such strong magnetic fields, photons undergo resonant scatterings by 
electrons in Landau levels to produce the Cyclotron Resonance Scattering Features (CRSF). 
These features have been found in more than 16 accretion powered pulsars \citep{miharaa,coburn2002}. 
The cyclotron line properties provide valuable information on the details of the 
emission, electron temperature, optical depth, and viewing geometries, and can map the 
magnetic field structure. For a brief review on the temporal and spectral
properties of transient Be/X-ray binary pulsars, refer \citet{paul2011}.

1A 1118-615 is a hard X-ray transient pulsar that was discovered with \emph{Ariel V} 
in an outburst in 1974 \citep {eyles1975}. The same series of observations 
revealed X-ray pulsations with a duration of 405.6 \rmfamily{s} \citep {ives1975}. The 
optical counterpart (He 3-640=``WRA 793'') is a highly reddened Be 
star classified as an O9.5IV-Ve star \citep {chevalier1975}. It shows strong Balmer 
emission lines indicating the presence of an extended envelope \citep{motch1988}. The distance to the source 
is estimated to be $5\pm2$ \rmfamily{kpc} from photometric and spectroscopic observations 
\citep {Janot1981}. \emph{UV} observations \citep {coe1985} reveal a P-Cygni profile in 
the C~IV line which indicates a stellar outflow with a velocity of the order of 1600$\pm$300 
km s$^{-1}$. \citet{coe1985} also confirmed the magnitude of the companion and from the extinction 
value, the distance was suggested to be about 4 \rmfamily{kpc}. An orbital period of 24$\pm$0.4 
days was reported by \citet {staubert2011}. There have been three giant outbursts so far 
detected for this source. The first was in 1974 \citep {marashi1976}. The second was detected 
in January 1992, and was observed by the Burst and Transient Source Experiment (\emph{BATSE})
covering an energy range of $15-50$ \rmfamily{keV}
 \citep{coe1994,marashi1976}. The outburst lasted 
for $\sim$30 days and pulsations with a period $\sim$406.5 s were detected up to 100 \rmfamily{keV} \citep{coe1994}. 
The same work also reported a spin change rate of -0.016 s/day at the decay of the outburst.

The source remained in quiescence for 17 years until 4 January 2009 when a third 
outburst was detected by \emph{Swift} \citep {mangano2009}. 
 Pulsations were detected at $\sim $ $407.68$ s with the pulse profile showing a complex structure. The 
complete outburst was regularly monitored with \emph{RXTE} (Rossi X-ray Timing Explorer). \emph{INTEGRAL} detected a flaring 
activity in the source $\sim30$ days after the main outburst \citep {leyder2009}. $Suzaku$ 
observed the source twice, once at the peak of the outburst and once 13 days later. 
\emph{RXTE} observations of the source was analyzed by \citet {doroshenko2011} who 
presented the broad-band spectrum, and reported the presence of a broad prominent absorption 
feature at $\sim$55 keV which they interpreted as a CRSF. Detailed timing and spectral analysis 
of the same set of observations was reported by \citet {devasia2011} which included the energy 
dependence of the pulse profiles and pulse phase-resolved spectroscopy of the continuum spectra. 
They also detected quasi-periodic oscillations (QPO) at 0.07-0.09 Hz which showed a significant 
energy dependence. Analysis of the same $Suzaku$ observations was reported by \citet {suchy2011} which 
confirmed the presence of the CRSF at $\sim$55 keV. They also performed pulse phase-resolved 
spectroscopy for three different pulse phases \emph{i.e.} the two peaks and the minimum of the 
pulse profile. \citet{suchy2011} thus reported a change in the continuum parameters like the photon index and high energy cutoff,
and also the centroid energy in the three above mentioned phases. 

 The pulse profiles of 1A 1118-615 show multiple energy dependent dips.
 Dips in the pulse profiles can be naturally produced by absorption in the accretion stream that is phase locked
with the neutron star. This had been explored 
for GX 1+4 \citep{galloway2001}, RX J0812.4-3114 \citep{corbet2000}, 1A 1118-61 \citep{devasia2011}, GX 304-1 \citep{devasia2011a} with \emph{RXTE},
 KS 1947+300 \citep{naik2006} with \emph{Beppo-Sax}, and GRO J1008-57
 with \emph{Suzaku} \citep{naik2011}. \citet{suchy2011} did include a phase dependent additional absorption component in their analysis but did not
 probe the narrow dips of the pulse profiles, \textbf{as} they carried out phase resolved analysis by dividing 
the pulse profile into only three phase bins. \citet {devasia2011} used \emph{RXTE-PCA} (\emph{RXTE} Proportional Counter Array) data covering a 
limited energy-band. We report here a more comprehensive timing and broad-band spectral analysis of \emph{Suzaku} 
observations of 1A~1118-61 during the 2009 outburst. We have probed the energy dependence of the 
pulse profiles in more detail and have carried out pulse phase-resolved spectroscopy in narrow 
phase bins to explain its complex nature.
\\We have also carried out pulse phase-resolved 
spectroscopy of the cyclotron absorption feature in this source and obtained one of 
the most detailed pulse phase-resolved information available for CRSF till date. In addition, 
we have detected a narrow phase dependent soft excess peak in both observations of this 
source. In section 2, we describe details of the observations and data reduction. In 
section 3, we describe the timing analysis including the energy dependence of the pulse profiles 
and the power density spectra. In section 4, we present the pulse phase averaged and pulse 
phase-resolved spectroscopy of the \emph{Suzaku} observations followed by discussions $\&$ 
conclusions in section 5.

\section[]{Observations \& Data reduction}

\begin{figure}
\includegraphics[scale=0.4,angle=-90]{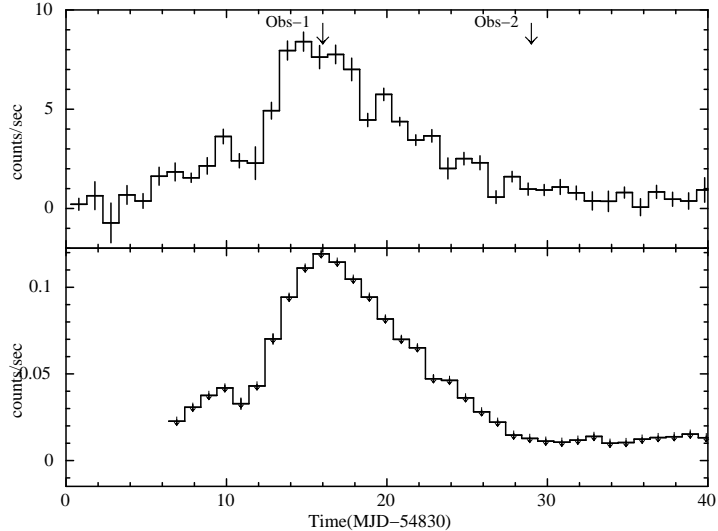}
\caption{The top panel shows the one day binned \emph{RXTE}-ASM (1.5-12 \rmfamily{keV}) light curve 
of 1A~1118-61, the bottom panel shows the one day binned light curve from the \emph{Swift}-BAT all 
sky monitor (15-50 \rmfamily{keV}). The arrows in the top panel indicate the time of the $Suzaku$ 
observations of 1A~1118-61.}
\label{fig1}
\end{figure}

The hard X-ray transient pulsar 1A~1118-61 was observed with \emph{Suzaku}, once at the 
peak of the outburst and once in its decline 13 days later.  Figure~\ref{fig1} shows the one 
day averaged long-term light curve of 1A~1118-61 using data from the the All Sky Monitor (ASM) ($1.5-12$ keV) onboard \emph{RXTE} and from the 
\emph{Swift}-BAT all sky monitor ($15-50$ keV) \footnote{See ftp://heasarc.gsfc.nasa.gov/xte/data/archive/ASMProducts/ and http://heasarc.nasa.gov/docs/swift/results/transients/ for details}. 
 The arrow marks in the figure shows
the $Suzaku$ observations of the pulsar during the 2009 outburst
\emph{Suzaku}
\citep {mitsuda2007}, covers the 0.2-600 \rmfamily{keV} energy range. 
It has two sets of instruments, the X-ray Imaging Spectrometer XIS \citep {koyama2007} covering the 0.2-12 
\rmfamily{keV} range, the Hard X-ray Detector (HXD) having PIN diodes \citep {takahashi2007} 
covering the energy range of 10--70 keV, and GSO crystal scintillators detectors covering the 70--600 keV
energy band. 
The XIS consists of four CCD detectors, three of them are front illuminated (FI) and one is back 
illuminated (BI).
\\ We have used the data processed with $Suzaku$ data processing pipeline 
ver. 2.3.12.25. The observations carried out on 15 January 2009 (Obs. ID--403049010;
hereafter quoted as Obs-1) has an useful exposure of 49.7 ks and that on 28 January 2009 (Obs. ID--403050010; 
hereafter quoted as Obs-2) has an useful exposure of 44.2 ks. In both observations, the source 
was observed at 'HXD nominal' pointing position \footnote{See http://heasarc.gsfc.nasa.gov/docs/suzaku/analysis/abc/
website for more information on instruments}. The XIS was operated in 'standard' data mode 
in the '$\frac{1}{4}$ window' option which gave a time resolution of 2 s. 

For extracting the XIS and HXD/PIN light curves and spectra, we used cleaned event data. 
The cleaned events of the XIS data were screened for standard criteria \footnote{See http://heasarc.nasa.gov/docs/suzaku/analysis/abc/node9.html for standard screeing criteria}.  We  
 extracted the XIS light curves 
and spectra from these cleaned events by selecting a circular region of $3^{'}$ around 
the source centroid. The XIS count rates for Obs-1 and Obs-2 were 44.5 c/s and 8.5 c/s, respectively. 
We also extracted the background light curves and spectra by selecting a region of the same size $18^{'}$ 
away from the source centroid. Response files and effective area files were generated by using the 
FTOOLS task 'xisresp'. For HXD/PIN background, simulated 'tuned' non X-ray background event files 
(NXB) corresponding to January 2009 were used to estimate the non X-ray 
background\footnote{http://heasarc.nasa.gov/docs/suzaku/analysis/pinbgd.html}, and the cosmic X-ray 
background was simulated as suggested by the instrument team\footnote{http://heasarc.nasa.gov/docs/suzaku/analysis/pin\_cxb.html} applying appropriate normalizations for both cases. 
Response files of the respective observations were obtained from 
the \emph{Suzaku} Guest Observer Facility.\footnote{http://heasarc.nasa.gov/docs/heasarc/caldb/suzaku/ and used for the HXD/PIN spectrum}

Observations of a bright source like 1A~1118-61, especially at the peak of the outburst should be 
checked for possible effects of pileup. \citet {suchy2011} had applied the tool 'pile\_estimate.sl' 
 to estimate the pileup fraction
in XIS and had excluded regions of the CCDs with pileup fraction above 5\%. 
We checked for the effect of pileup for Obs-1 by extracting spectra with the central 
$10^{''}$ and $15^{''}$ radii removed, and generated their corresponding responses. Comparing these two spectra with the original spectrum, we observe that the spectral 
shape remains almost the same in all the cases and the change in photon index is only $\sim$0.02 for the XISs, where this  
effect is the most pronounced for XIS-3. Hence, our results are not expected to vary significantly due to 
this effect, since our main focus is on variations of the spectral parameters with pulse phase that is much
 larger than this difference. Obs-2, being an order of magnitude fainter than Obs-1 does not require this check.


\section{Timing Analysis}

\subsection{Pulse profiles}
\begin{figure}
\includegraphics[height=8 cm]{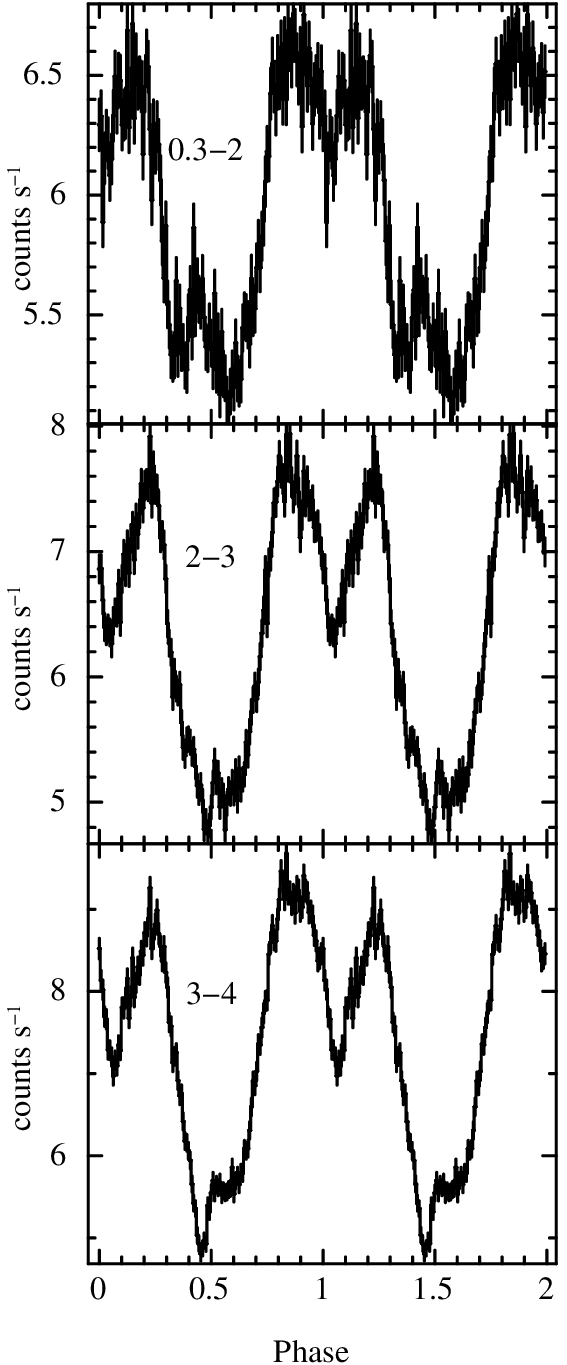}
\includegraphics[height=8.2 cm]{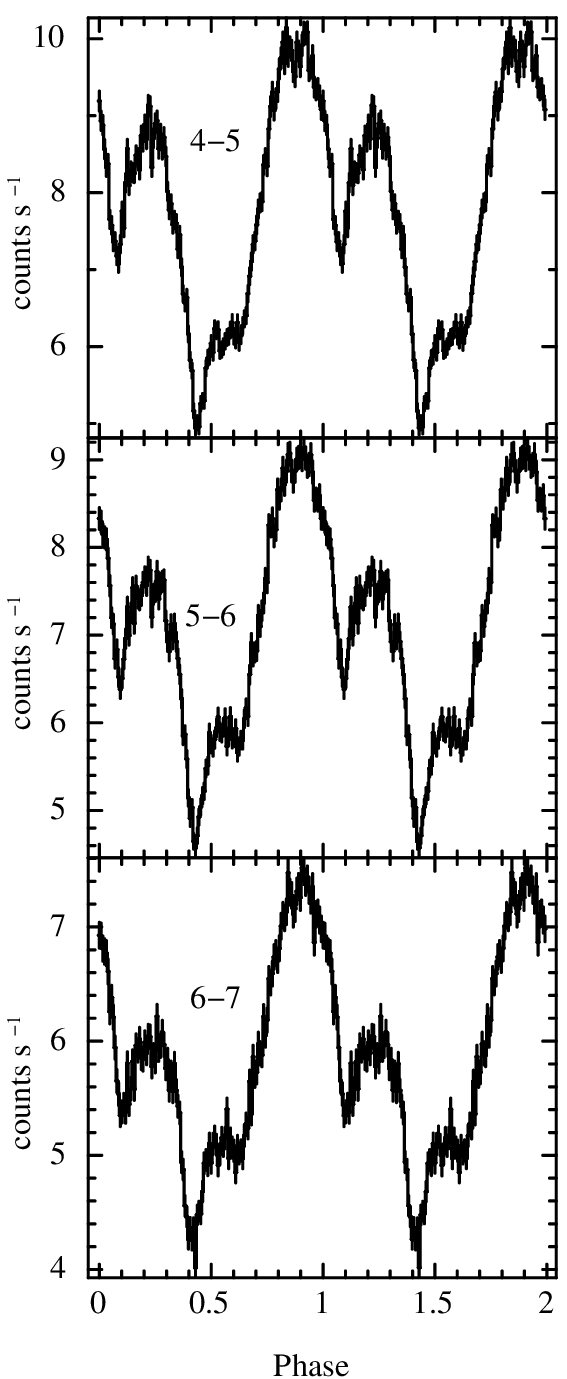}
\includegraphics[height=8 cm]{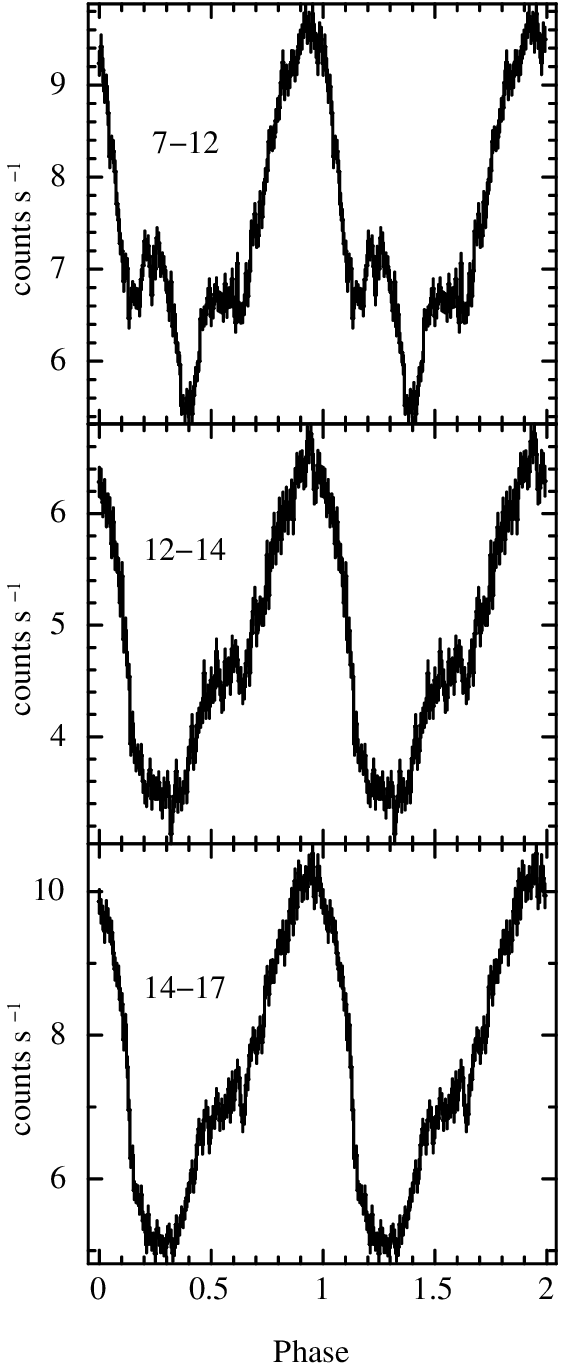}
\includegraphics[height=8 cm]{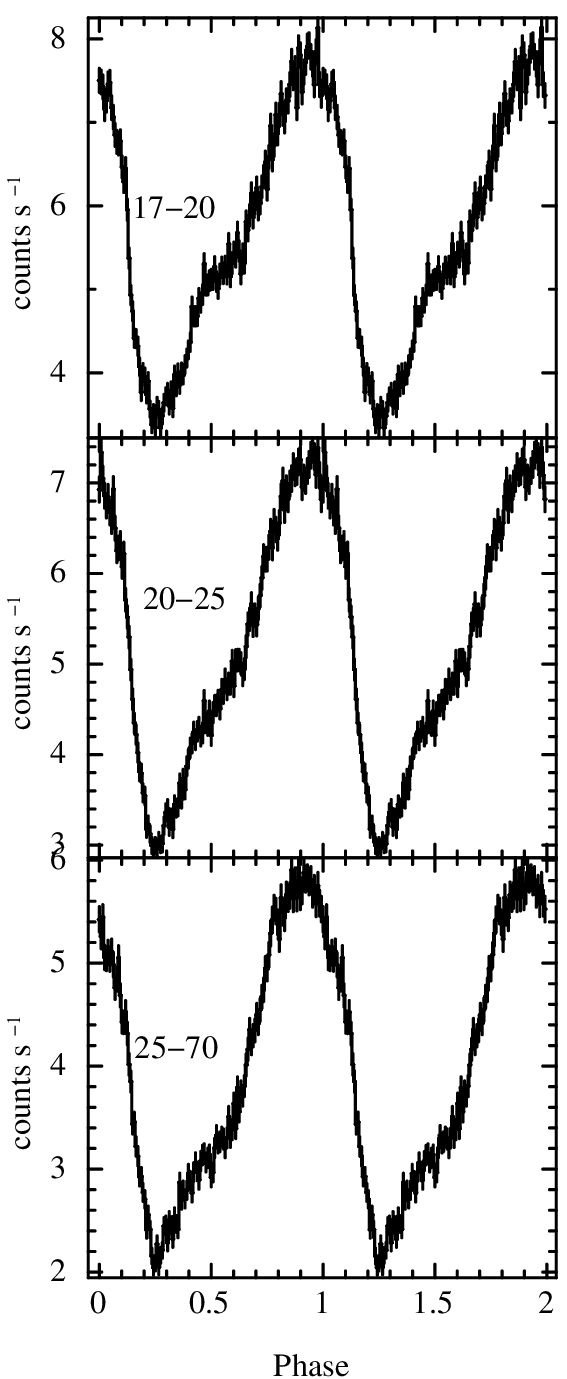}
\caption{Energy dependent pulse profiles of 1A~1118-61 using the XIS and PIN data for Obs-1. 
Captions inside the figure (in units of \rmfamily{keV}) represent the energy range for which the profiles were created. The 
pulse profiles in the 0.2--12 keV energy range are made from the XIS data (data from all the three XISs are added together), 
where as profiles in the 12--70 keV range are obtained from the PIN data.}
\label{fig2}
\end{figure}

\begin{figure}
\includegraphics[height=8 cm]{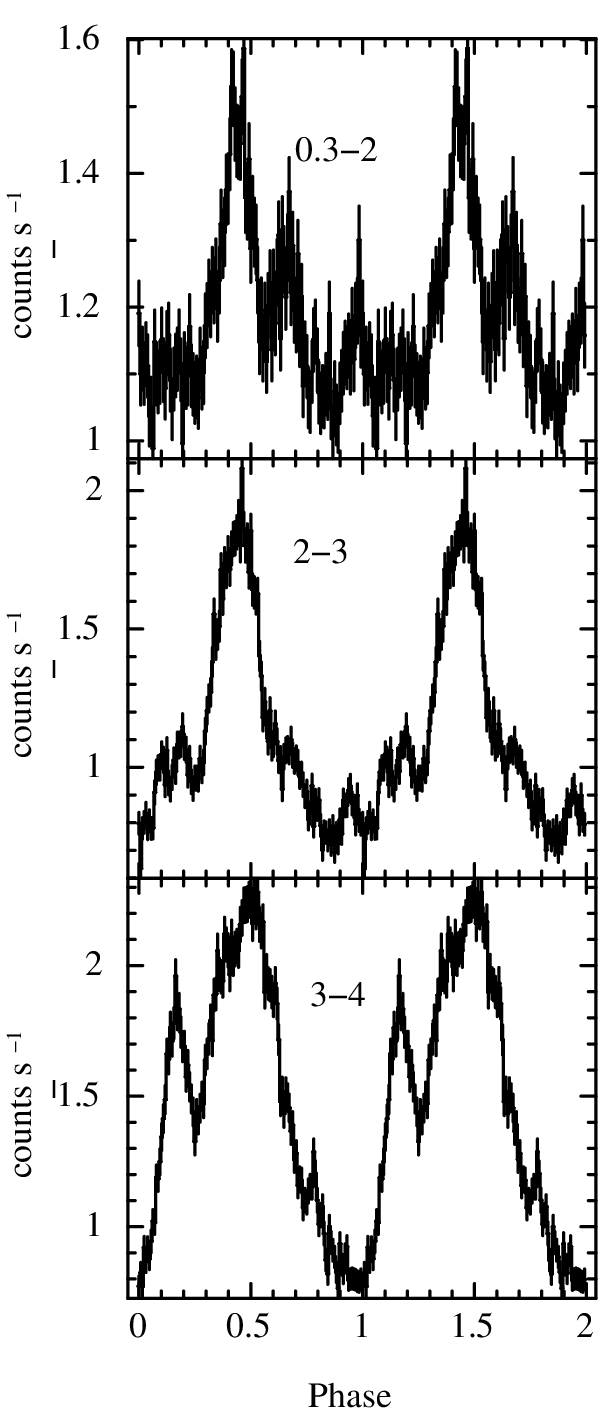}
\includegraphics[height=8 cm]{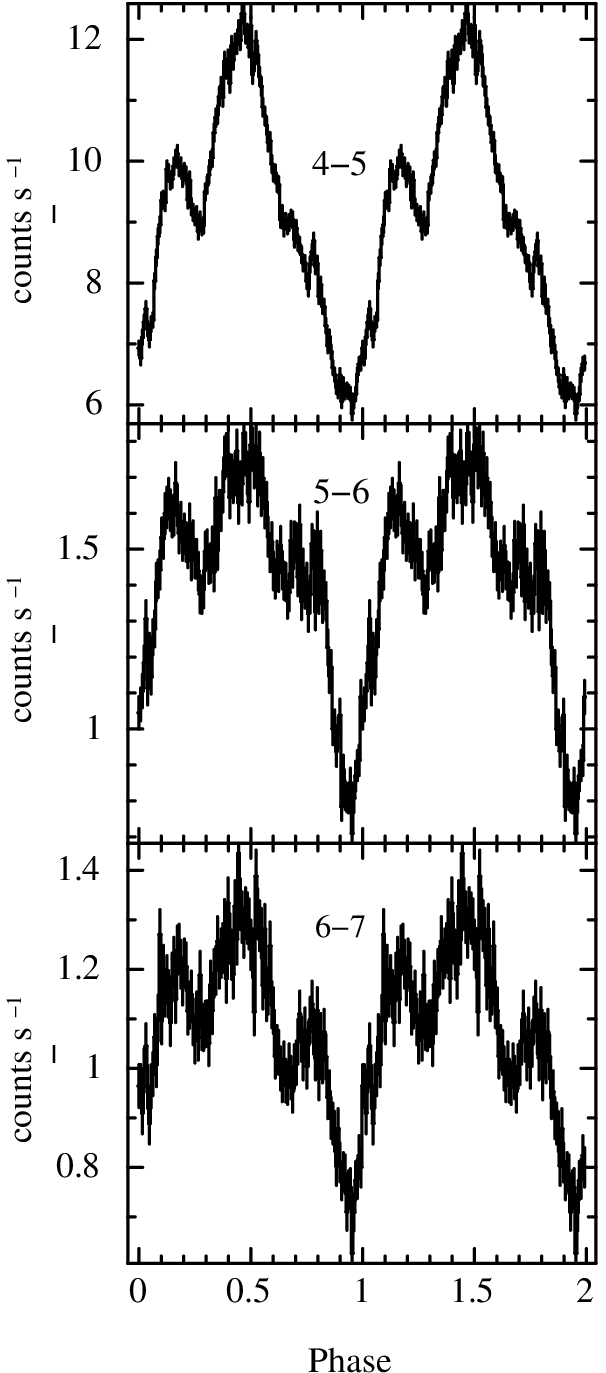}
\includegraphics[height=8 cm]{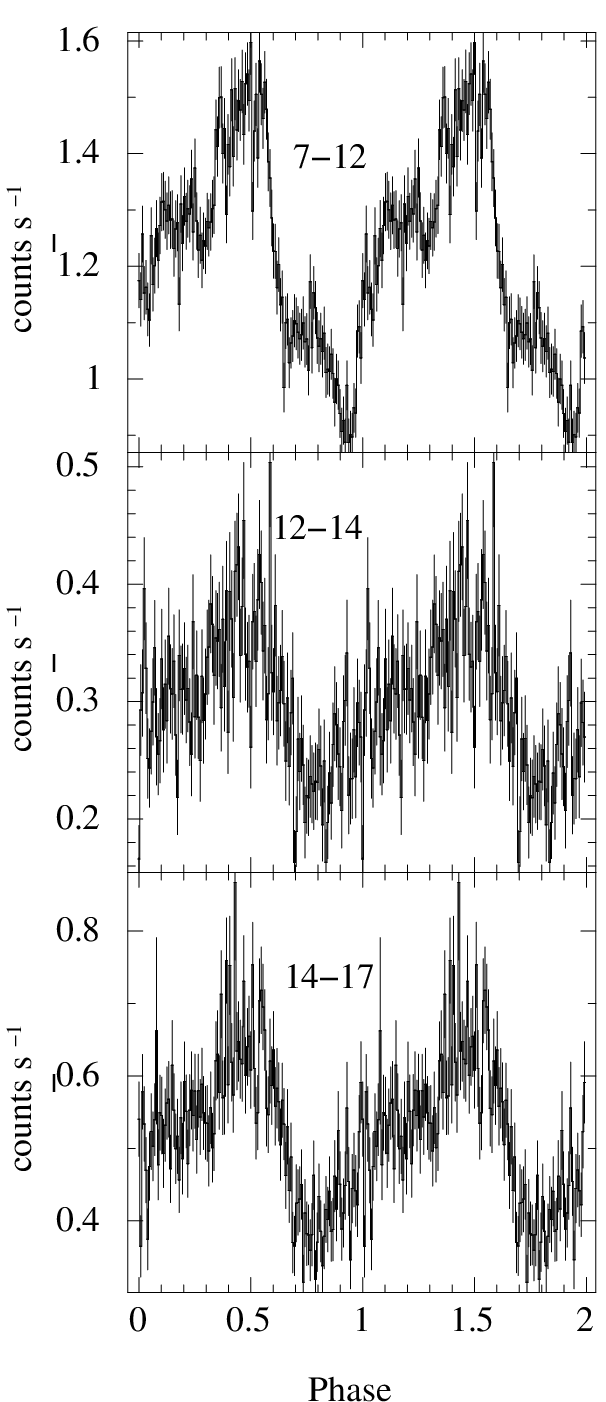}
\includegraphics[height=8 cm]{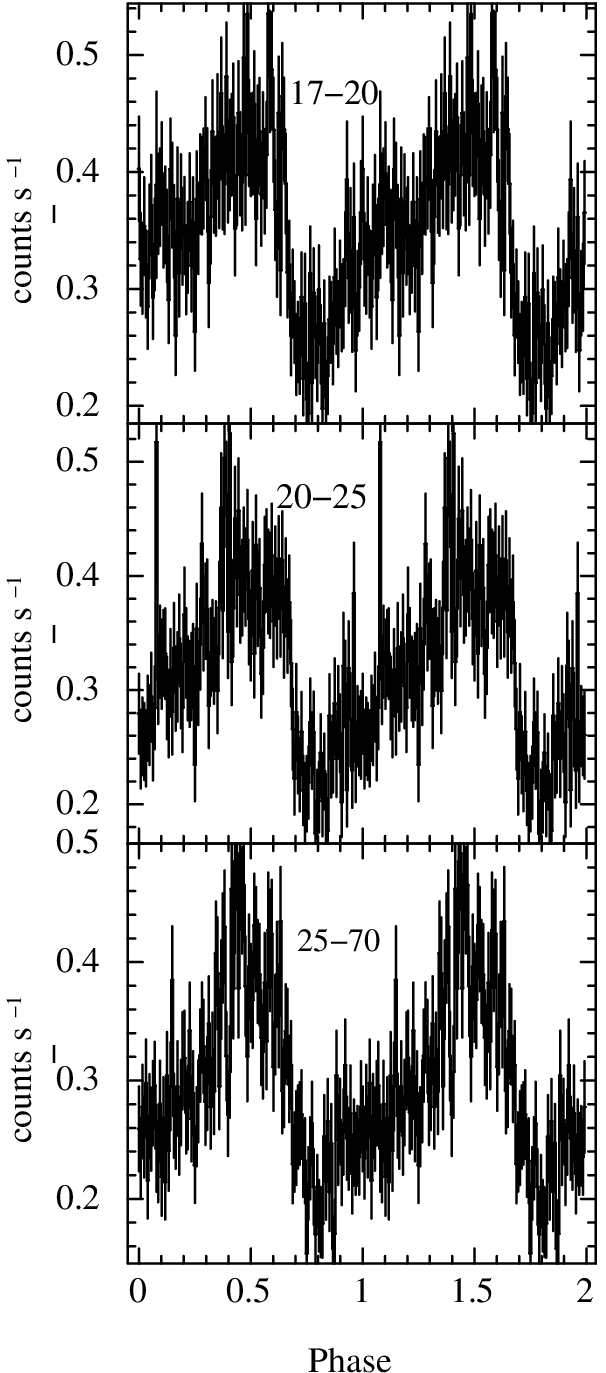}
\caption{Same as Figure 2 for Obs-2.}
\label{fig3}
\end{figure}

For the timing analysis, we applied barycentric corrections to the event data files using the FTOOLS 
task 'aebarycen'. Light curves with a time resolution of 2 s and 1 s were extracted from the XIS (0.2--12 keV) 
and the HXD/PIN (10--70 keV), respectively. We searched for pulsations in both observations by applying 
pulse folding and $\chi^{2}$ maximization technique. The pulse period was found to be 407.49$\pm$0.73 s 
for Obs-1 (MJD 54846) and 407.40$\pm$0.40 s for Obs-2 (MJD 54859) respectively.
 We created energy resolved pulse profiles by folding the light curves in different energy bands with 
the respective pulse periods. Light curves from the three XISs were added together to create the 
pulse profiles in the 0.2--12 keV energy range. The pulse profiles are found to have complex energy dependent structures on both the days, with 
more than one peak in the XIS energy band, and a single peaked but asymmetric structure in the PIN energy band, 
as shown in Figures~\ref{fig2} \& \ref{fig3}. 
In Obs-1 (Figure \ref{fig2}), the profiles show a double peaked structure at lower energies (less than 12 \rmfamily{keV}). The secondary peak (phase $\sim$ 0.2) decreases  
in strength with energy and disappears at $\sim$12 \rmfamily{keV}. At high energies (greater than 12 \rmfamily{keV}), the profile shows a single peaked 
structure with a large pulse fraction. A narrow low energy peak (less than 2 \rmfamily{keV}) is seen coincident with the pulse 
minima at higher energies (phase $\sim$ 0.46). For Obs-2 (Figure \ref{fig3}), one main peak (phase $\sim$ 0.5) is seen at all energy bands, which is narrow and 
symmetric at low energies (less than 3 \rmfamily{keV}). The main peak gets broadened up to 12 keV, and becomes a broad asymmetric pulse at higher energies. A narrow and
weaker peak (at phase $\sim$ 0.2) appears at energies greater than 2 \rmfamily{keV} before the main peak. At higher energies (greater than
12 \rmfamily{keV}), this feature gradually merges with the rising 
part of the main peak. Another pulse component (phase $\sim$ 0.75) is seen after the main peak only in the energy band below 2 \rmfamily{keV}
and 6--7 \rmfamily{keV}. The pulse profiles show a single asymmetric peaked structure in the energy ranges greater than 12 \rmfamily{keV} with the pulse minima 
at phase $\sim$ 0.8. The pulse profiles show the presence of an additional peak at phase $\sim$ 0.96 in the lowest energy bin (0.3--2 keV). The complex energy dependence of the pulse profile implies significant changes in the energy
 spectrum with the pulse phase, and also a complex underlying spectrum. A detailed discussion is presented in section 5.1.

\subsection{Power Density Spectra}

\begin{figure}
\includegraphics[scale=0.35,angle=-90]{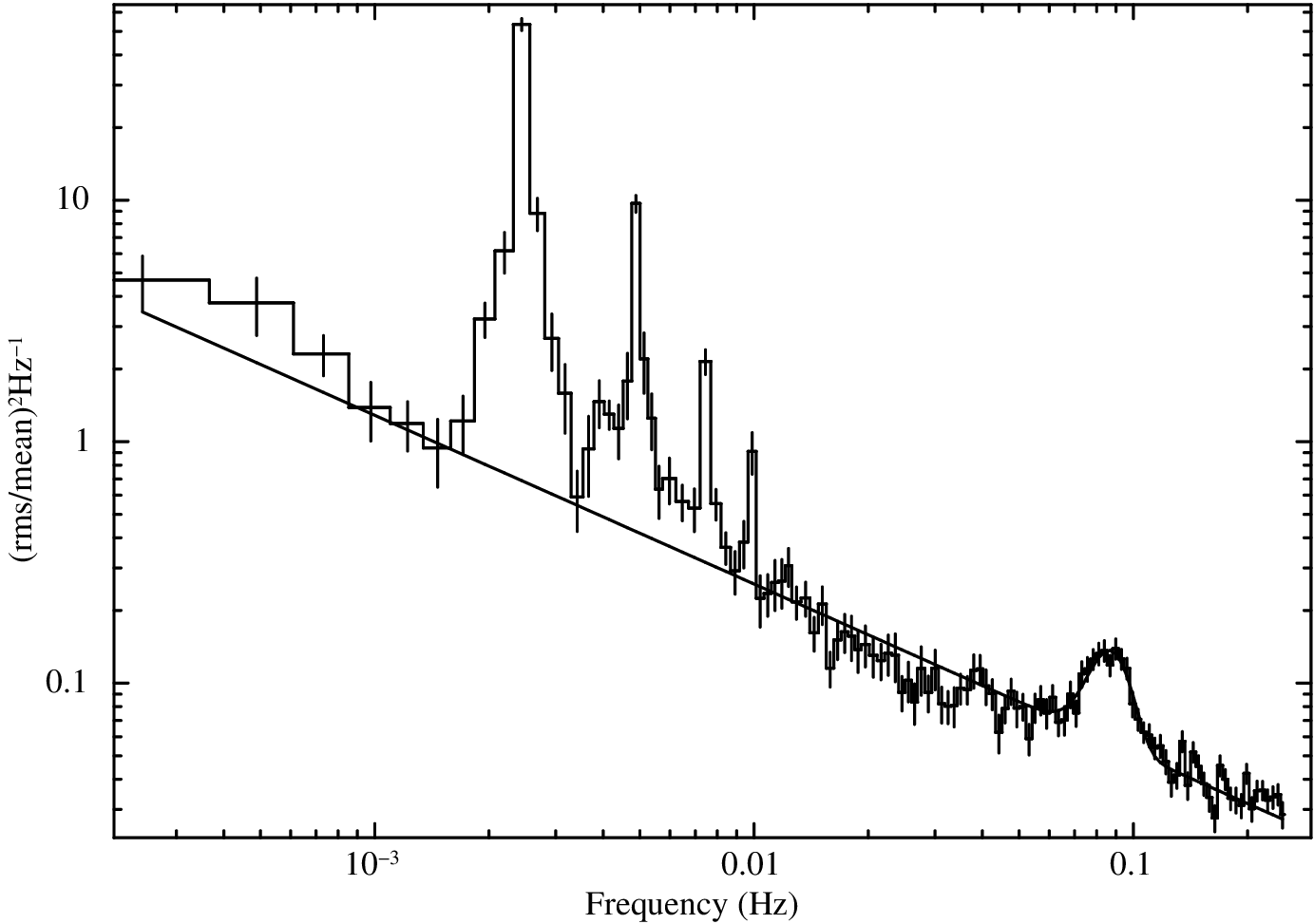}
\includegraphics[scale=0.35,angle=-90]{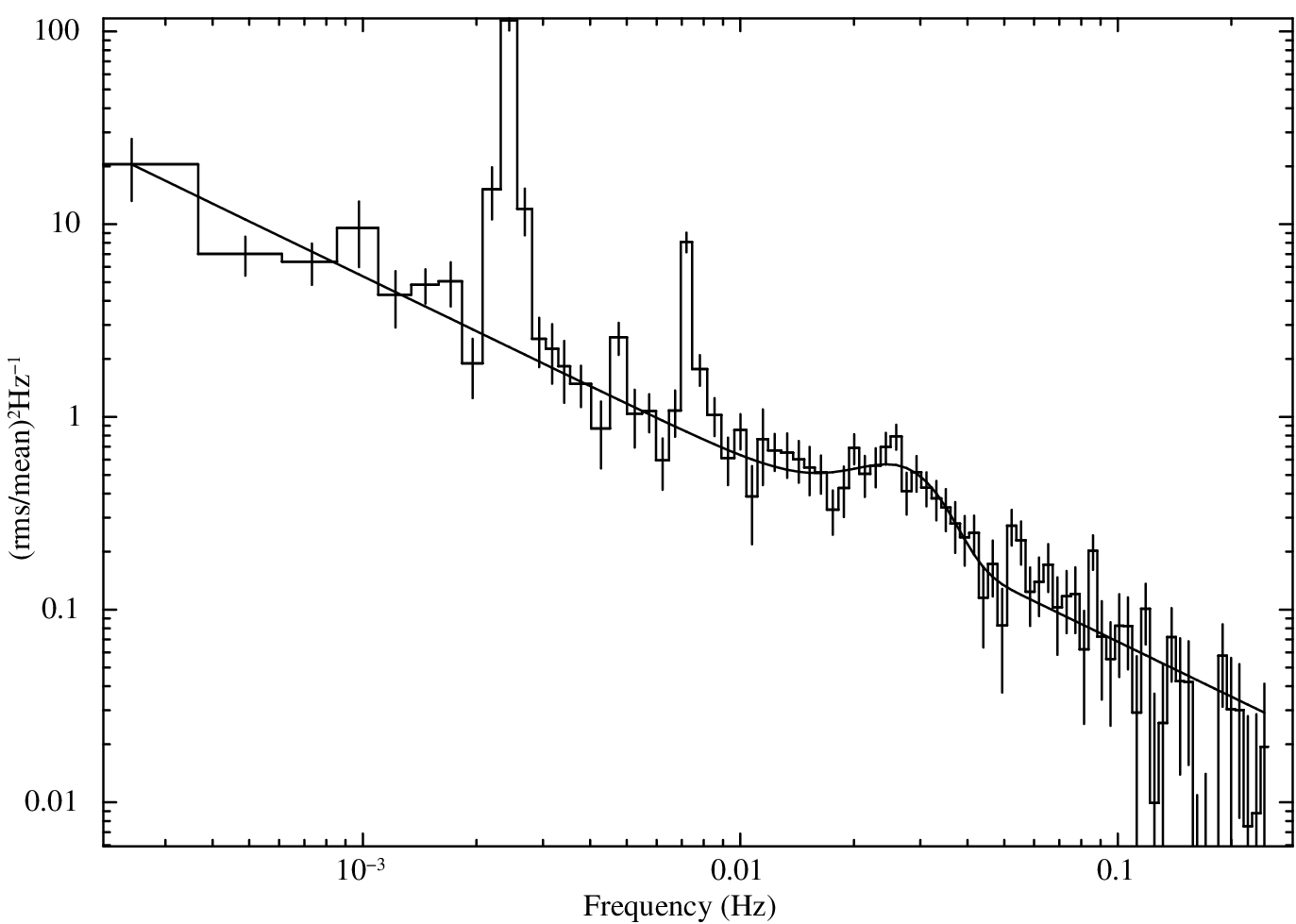}
\caption{Average PDS of 1A~1118-61 for both observations showing the QPO feature in the 
energy band of 0.2--12 \rmfamily{keV}. The left panel shows the PDS for Obs-1 and the right panel shows 
the same for Obs-2. In both the figures, the spin period at 0.0025 Hz and its harmonics are clearly 
seen with the QPO at 0.087 Hz visible for Obs-1 and the broad feature at 0.026 Hz visible for Obs-2.}
\label{fig4}
\end{figure}

We created the Power Density Specta (PDS) using the FTOOL task '\emph{powspec}' 
from the XIS light curves (0.2-12 \rmfamily{keV}) which was divided into segments of 4096 s 
with a time resolution 2 s. The PDS obtained from all these segments were rebinned with a 
geometric rebinning factor, and were averaged to produce the final PDS. The average PDS was also normalized such that its integral gave the squared rms fractional 
variability and expected white noise level was subtracted. The PDS from both observations clearly 
shows the peak at $\sim$0.0025 Hz corresponding to the spin frequency, with its harmonics
visible at higher
frequencies. Obs-1 
shows a quasi-periodic oscillation (QPO) feature at $0.087\pm0.008$ Hz. 
\\ \
We fitted the PDS with a power-law 
component corresponding to the continuum by avoiding the peaks corresponding to the spin 
frequency and its harmonics. The QPO feature was fitted with a Gaussian function, giving a detection significance of 13$\sigma$, quality factor ($\nu$/\emph{FMHM)} of $4.2\pm0.4$ and the rms 
fractional variability of $4.58\pm0.17$ $\%$ in the 0.2--12 \rmfamily{keV} energy range of XIS.
Obs-2 shows a broad feature at $0.026\pm0.003$ Hz. We have fitted the feature with a Gaussian function which gave a 
detection significance of $7\sigma$, quality factor of $1.70\pm0.2$ and rms value of $7.93\pm 0.58 \%$. 
The energy dependence of the QPO has been investigated for Obs-1. Multiple PDS were extracted 
for seven different energy ranges between 3 and 10 \rmfamily{keV}, each with a width of 1 \rmfamily{keV} in the XIS 
energy band. The rms of the QPO feature shows an increase from 4.09$\%$ to 7.74$\%$ in the XIS energy 
range. Figure~\ref{fig4} shows the PDS for both observations in the energy band of 0.2--12 keV.
We note here that it is quite unusual to detect QPOs with imaging spectrometer detectors. 
In this source we have not only detected the QPOs, but are also able to investigate it's energy dependence.


\section{spectral analysis}

\subsection{Pulse phase averaged spectroscopy} 

The continuum emission from HMXB accretion powered pulsars can be interpreted as Comptonization of 
 soft X-rays in the plasma above the neutron star surface. It can be modelled as a power-law with an 
exponential cutoff or as a broken power-law \citep {white1983, miharaa, coburn2002}.

In the presence of strong magnetic fields, the 
continuum photons produced inside the hot region by thermal bremsstrahlung and Comptonization may get 
resonantly scattered in the line forming regions, producing a cyclotron resonant scattering feature 
(CRSF). Often a strong narrow Fe $K\alpha$ line can also be found originating from the cold dense 
matter in the vicinity of the neutron star. 

\begin{figure}
\includegraphics[scale=0.35,angle=-90]{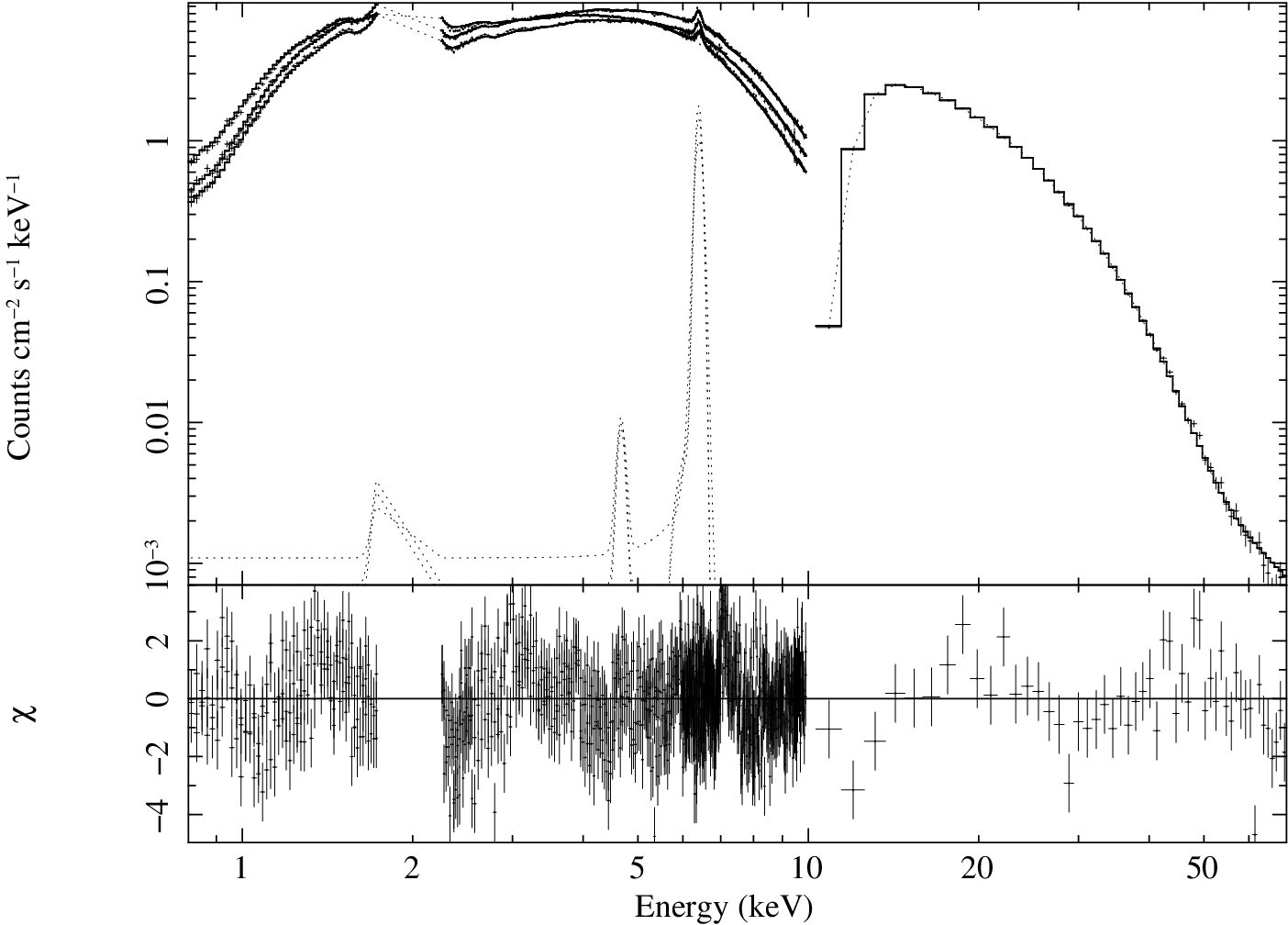}
\includegraphics[scale=0.35,angle=-90]{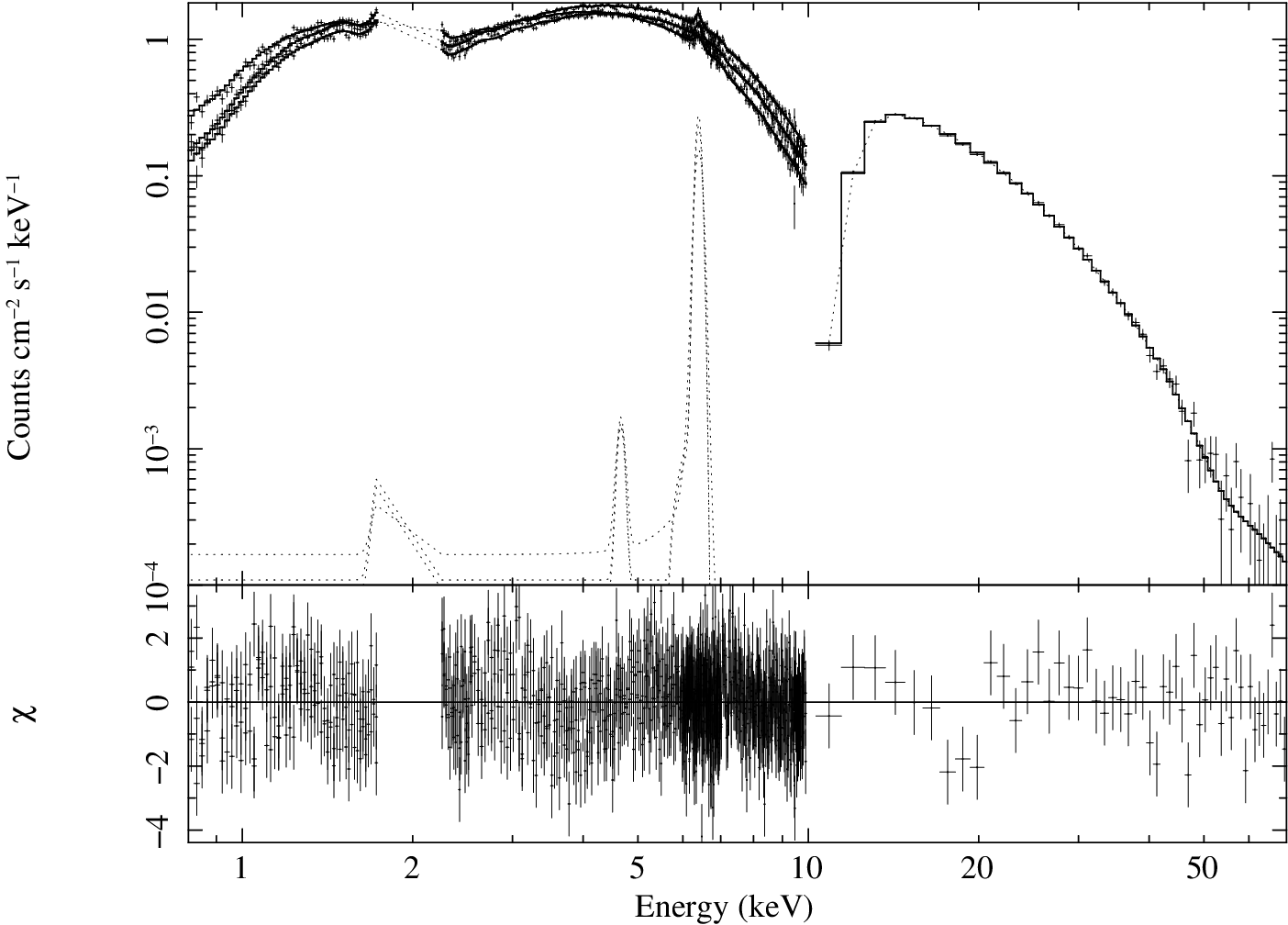}
\caption{The pulse-phase averaged spectrum of 1A~1118-61 at the peak of the outburst (Obs-1; left panel) 
and at the decline (Obs-2; right panel), using XIS from 0.8--10 \rmfamily{keV} and PIN from 10--70 \rmfamily{keV}. 
The upper panel shows the best-fit spectra along with the model components. The residuals are given in \textbf{the} bottom 
panels.}
\label{fig6}
\end{figure}

 We  performed pulse phase averaged spectral analysis of 1A~1118-61 using spectra from the front illuminated 
CCDs (XISs-0 and 3), back illuminated CCD (XIS-1), and the PIN. Spectral fitting was performed using \emph{XSPEC} v12.6.0. The energy ranges 
chosen for the fits were 0.8--10 \rmfamily{keV} for the XISs, and 10--70 \rmfamily{keV} for the PIN spectra. 
We neglected the 1.75--2.23 \rmfamily{keV} energy range in order to take out the artificial structures in the XIS spectra at around the Si edge and Au edge. After appropriate background subtraction we fitted the  spectra simultaneously with all the parameters tied, except the relative instrument normalizations which
were kept free. The XIS spectra were rebinned by a factor of 5 from 0.8--6 \rmfamily{keV} and 7--10 \rmfamily{kev}. However due to the presence of a narrow iron line, the spectrum was rebinned only 
by a factor of 2 in the 6--7 \rmfamily{keV} energy range. The PIN spectra were rebinned by a factor of 3 in the 10--70 \rmfamily{keV} range.
We had initially tried to 
fit the continuum with a partial covering high-energy cutoff model (similar modeling for this source has been 
done in \cite {devasia2011}) which gave a very low value of high-energy cutoff ($\sim$0--3 \rmfamily{keV}) for some of the phase 
bins while performing pulse phase resolved spectroscopy described later in section 4.2. High energy cutoff in this
scenario is equivalent to the cutoff power-law model.
Hence we fitted the broad band energy spectrum of 1A~1118-61 with a partial covering cutoff power-law model with an 
interstellar absorption and a gaussian line for an Fe $K\alpha$ feature at 6.4 \rmfamily{keV}. In addition, a broad cyclotron absorption feature was 
also found in the spectra at $49 \pm 0.5$ \rmfamily{keV} in Obs-1 and  $52^{+4.9}_{-3.1} $ \rmfamily{keV}
in Obs-2. This feature was fitted with the \emph{XSPEC} standard model 'cyclabs' which 
is a cyclotron absorption line having a Lorentzian profile. Presence of the broad cyclotron feature at 
$\sim55 $ \rmfamily{keV} had already been reported from \emph{RXTE} observations \citep {doroshenko2011} and 
from these \emph{Suzaku} observations \citep {suchy2011}. The analytical form of the model which was used to 
fit the spectra is as follows:
\begin{equation}
 \mathrm{F(E)}=\exp^{-\sigma(\mathrm{E})\mathrm{N_{H1}}}((\mathrm{N}1+\mathrm{N}2~e^{-\sigma(E)\mathrm{N}_{\mathrm{H}2}})*CYABS(E)+G)~I(E)
\end{equation}
  Where CYABS is the Lorentzian cyclotron line profile given by
\begin{equation}
 CYABS(E)=\exp[-D_{Cycl}\frac{({W_{Cycl}E/E_{Cycl}})^{2}}{{(E-E_{Cycl})^{2}+{W_{Cycl}}^{2}}}]
\end{equation}
where $E_{Cycl}$ is the energy of the cyclotron line \& $D_{Cycl}$ \& $W_{Cycl}$ are the depth and width of the cyclotron line. I(E) can be expressed as 
\begin{equation}
 \mathrm{I(E)}=E^{-\Gamma}\exp({\frac{-E}{\beta})}
\end{equation}
Where $\Gamma$ is the power-law photon index and $\beta$ is the e-folding energy of the exponential roll-over 
in \rmfamily{keV}. N1 and N2 are the normalizations of the two power-laws. $N_{H1}$ is the Galactic hydrogen 
column density along our line of sight, and $N_{H2}$ is the column density of the material that is local to 
the neutron star. $\sigma$ is the photo-electric cross-section and G is the gaussian line energy. Thus the 
covering fraction of the more absorbed power-law is given by $Cv_{fract}=\frac{\mathrm{N2}}{\mathrm{N1}+\mathrm{N2}}$. 
\\ \ Spectral fitting showed that Obs-1 was softer than Obs-2. Since the energy spectrum had a strong pulse 
phase dependence, we did not expect the pulse averaged spectra to fit well to a simple continuum model. 
The best-fit had a reduced $\chi^2$ of 1.79 for 1022 degrees of freedom and 1.21 for 1023 
degrees of freedom for Obs-1 and Obs-2, respectively. Table~1 shows the best-fit parameters for the model. 
Figure~\ref{fig6} shows the best-fit spectra along with the residuals for both observations.

\begin{table*} 
\caption{Best fitting spectral parameters of 1A~1118-61. Errors quoted are for 90 per cent confidence range.}
\label{table}
\centering
\begin{tabular}{@{}l c c@{}}
\hline \hline
Parameter &  15 January (Obs-1) & 28 January (Obs-2) \\
\hline
$N_{H1}$ ($10^{22}$ atoms $cm^{-2}$) & $0.86 \pm 0.01$ & $0.61_{-0.02}^{+0.01}$\\
$N_{H2}$ ($10^{22}$ atoms $cm^{-2}$)  & $14.25_{-0.59}^{+0.71}$ & $7.78_{-0.24}^{+0.14}$ \\
$Cv_{fract}$ & $0.24 \pm 0.01$ & $0.73 \pm 0.01$ \\
PowIndex ($\Gamma $) & $0.37 \pm 0.02$ & $0.97_{-0.04}^{+0.02}$\\
E-folding energy $ \beta$ (\rmfamily{keV}) & $21.6_{-0.6}^{+0.7}$ & $24.8_{-2.4}^{+1.6}$ \\
$E_{Cycl}$ & $49.3_{-0.5}^{+0.5}$ & $51.9_{-3.1}^{+4.9}$ \\
$D_{Cycl}$ & $1.70 \pm 0.05$ & $1.00_{-0.3}^{+0.5}$ \\
$W_{Cycl}$ & $19.2 \pm 1.1$ & $10.7_{-2.5}^{+3.5}$ \\
Iron line energy (\rmfamily{keV}) & 6.40$\pm$ 0.01 & 6.40$ \pm $0.01 \\
Iron line eqwidth (\rmfamily{eV}) & 51.0 $\pm $1.5 & 41.9 $\pm $4.0\\
Flux (XIS) (0.8-10 \rmfamily{keV}) $^a$ &  2.45 $\pm$ 0.01  &  0.45 $\pm$  0.04 \\
Flux (PIN) (10-70 \rmfamily{keV}) $^b$ &  1.02 $\pm$ 0.01 & 0.11 $\pm$ 0.04     \\
reduced $\chi^{2}$/d.o.f & 1.79/1022 & 1.21/1023 \\
\hline
\end{tabular}\\
$^a$ Flux is in units of $10^{-9}$ ergs $cm^{-2}s^{-1}$\\
$^b$ Flux is in units of $10^{-8}$ ergs $cm^{-2}s^{-1}$
\end{table*}

\subsection{Pulse phase-resolved spectroscopy}
\begin{figure}
\includegraphics[scale=0.35,angle=-90]{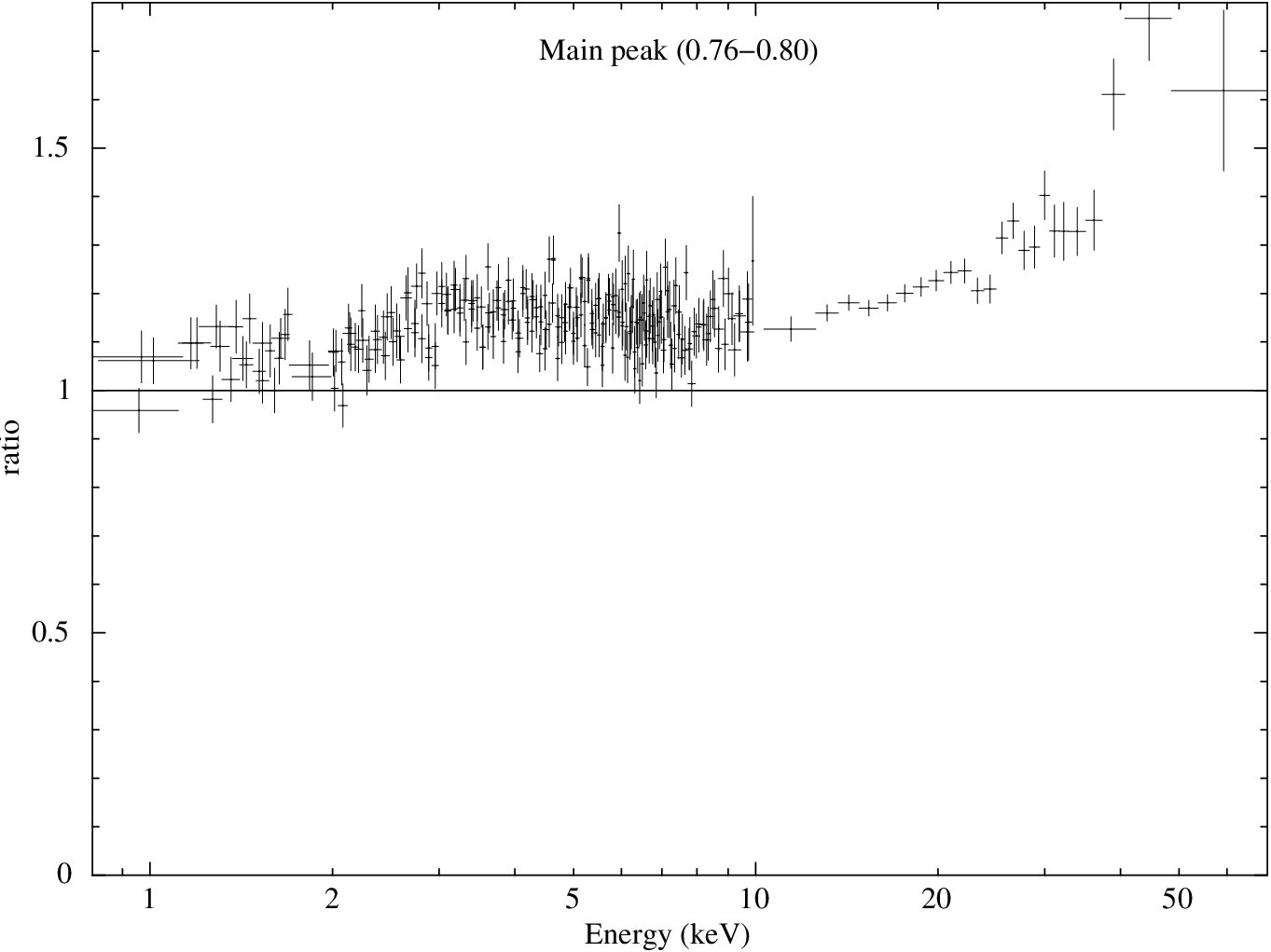}
\includegraphics[scale=0.35,angle=-90]{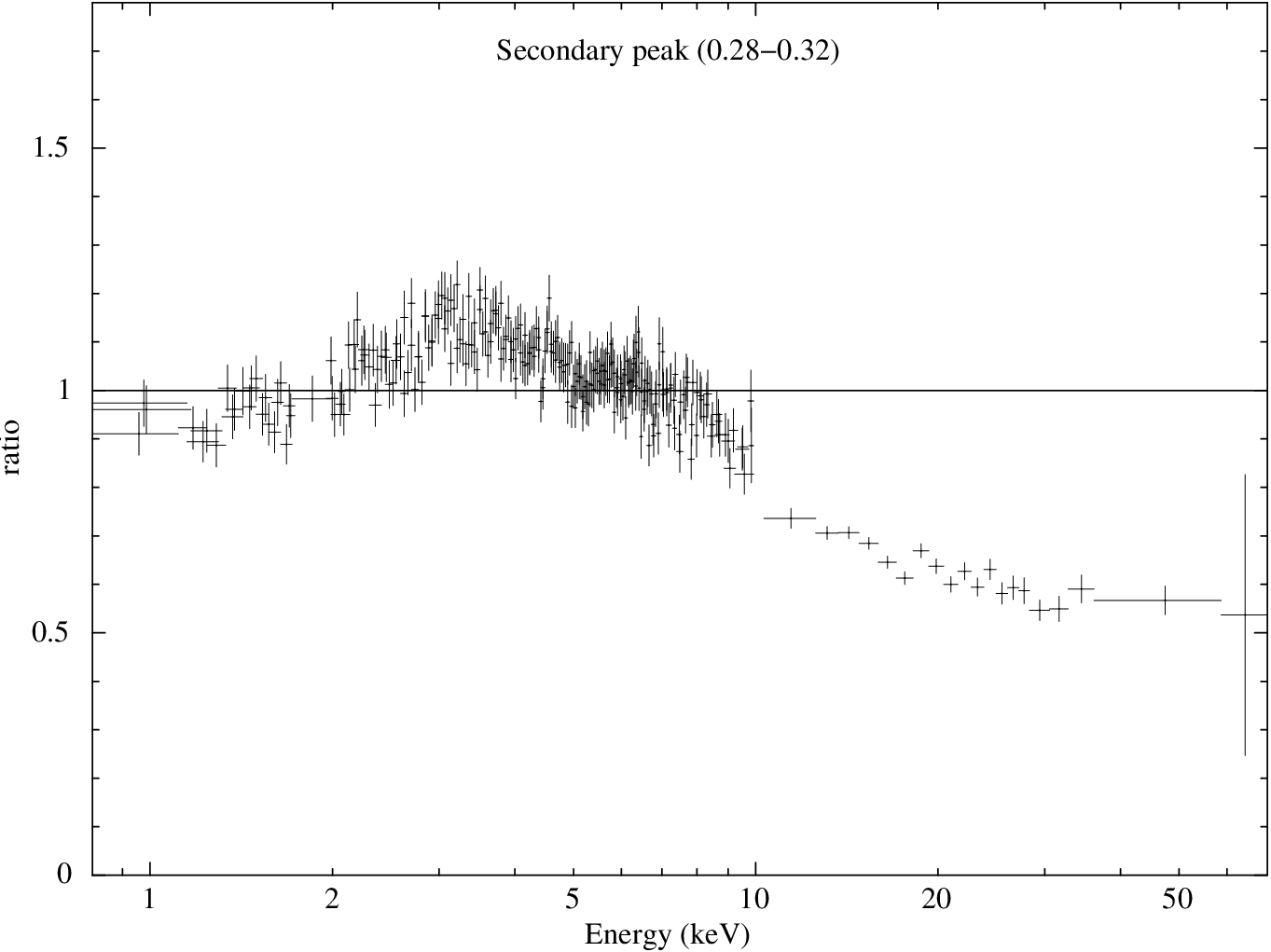}
\includegraphics[scale=0.35,angle=-90]{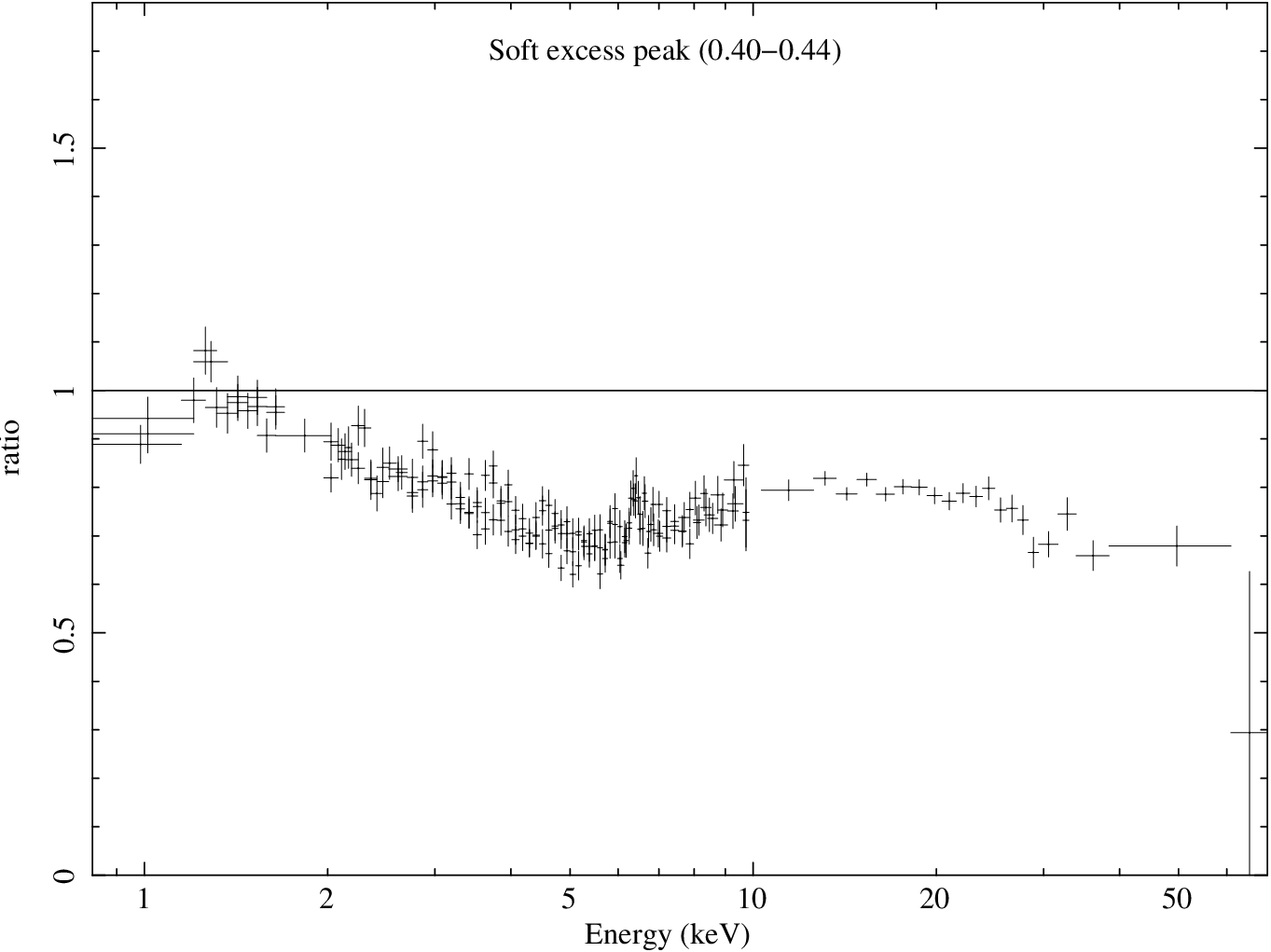}
\includegraphics[scale=0.35,angle=-90]{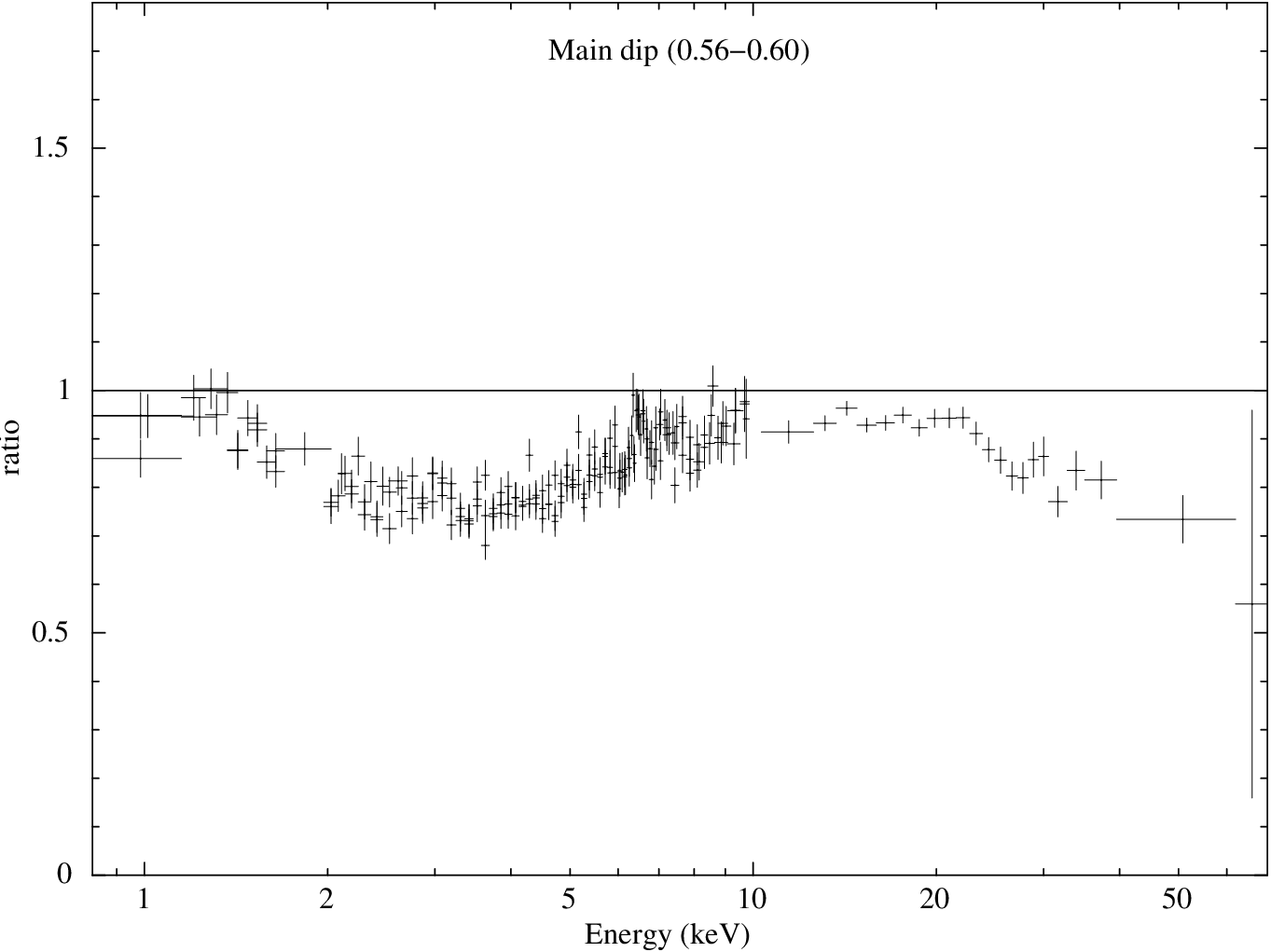}
\caption{Phase-resolved to phase averaged spectral ratios for some particular phase bins 
(see values in parentheses) at the peak of the outburst on 15 January 2009 (Obs-1).}
\label{fig7}
\end{figure}

\begin{figure}
\includegraphics[scale=0.35,angle=-90]{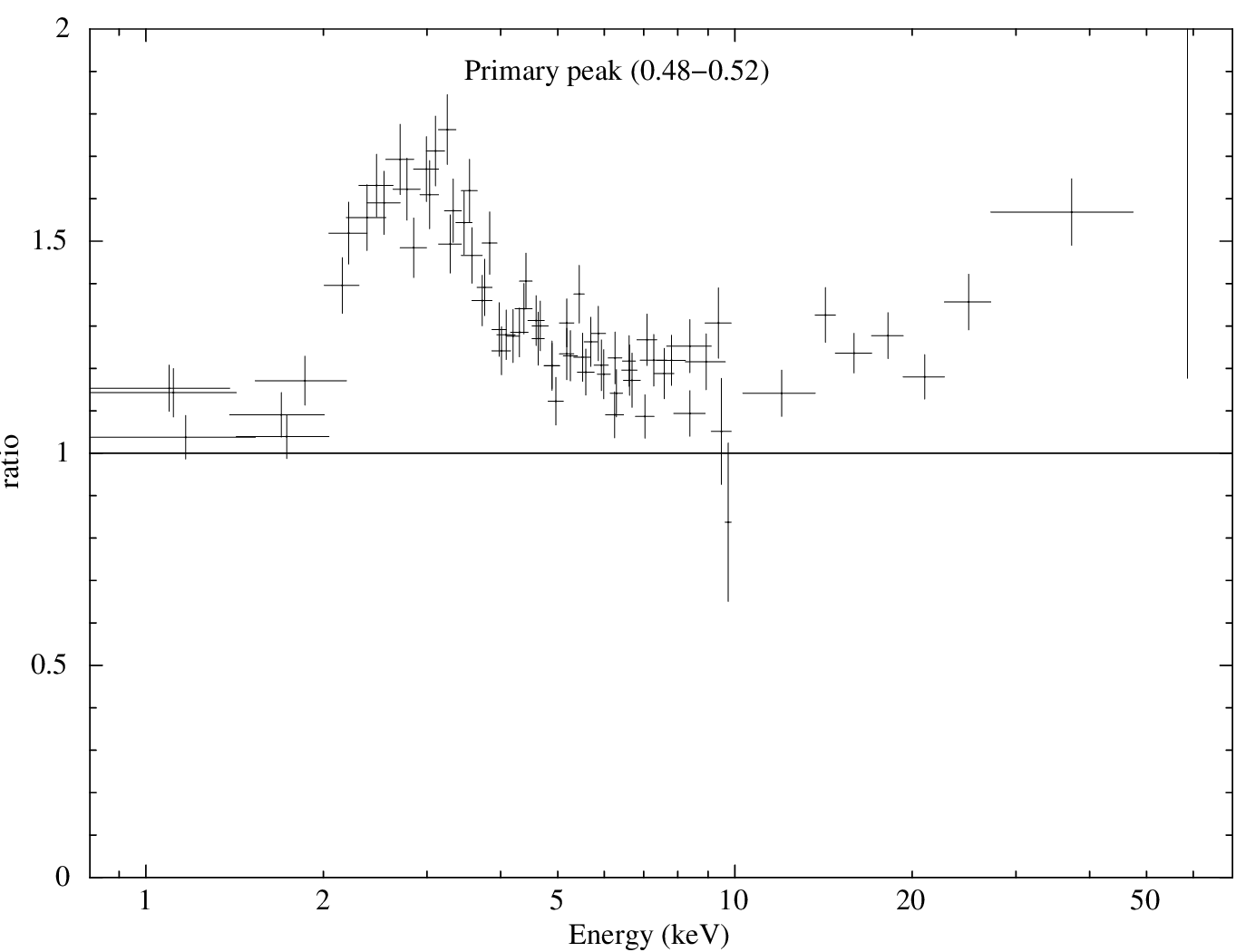}
\includegraphics[scale=0.35,angle=-90]{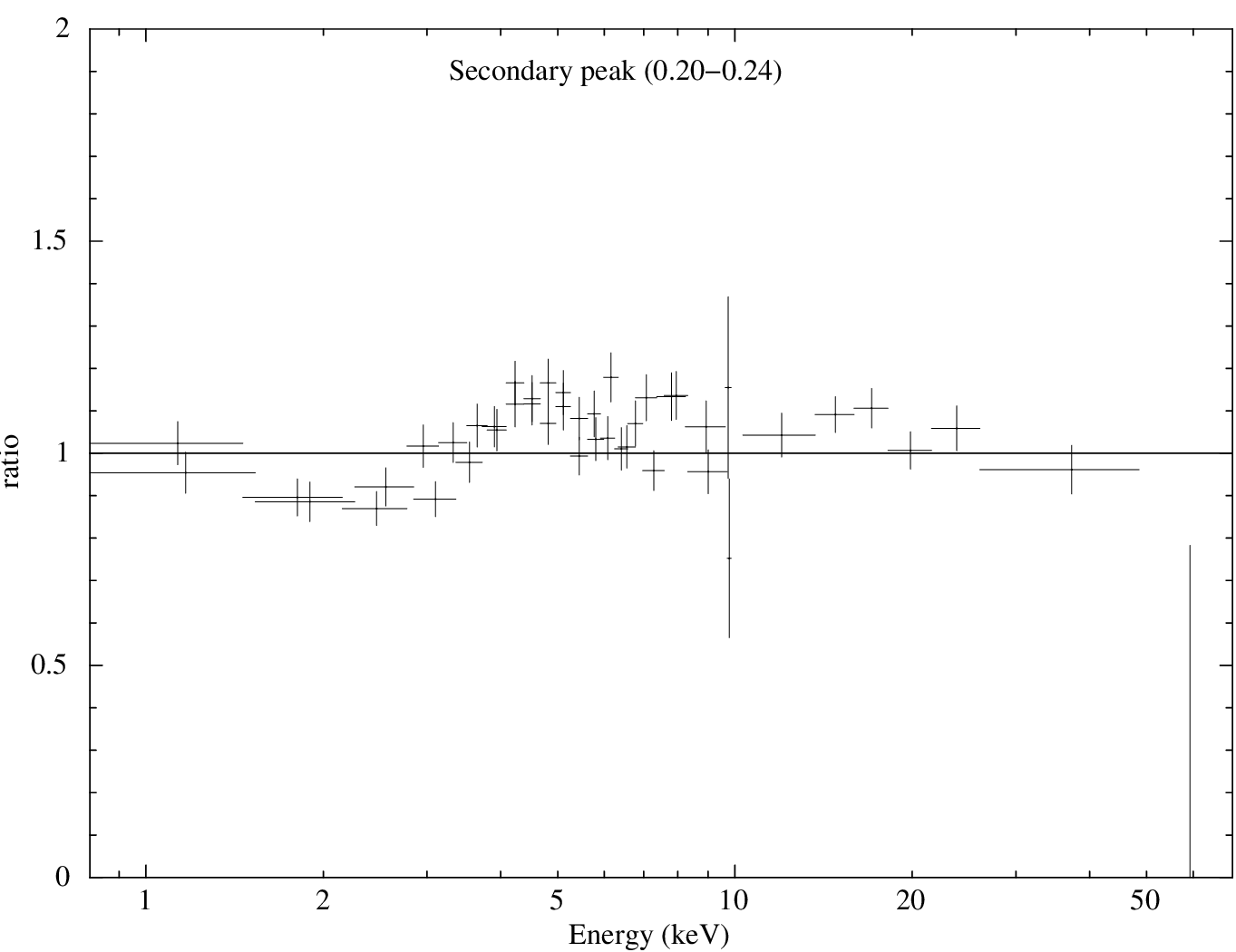}
\includegraphics[scale=0.35,angle=-90]{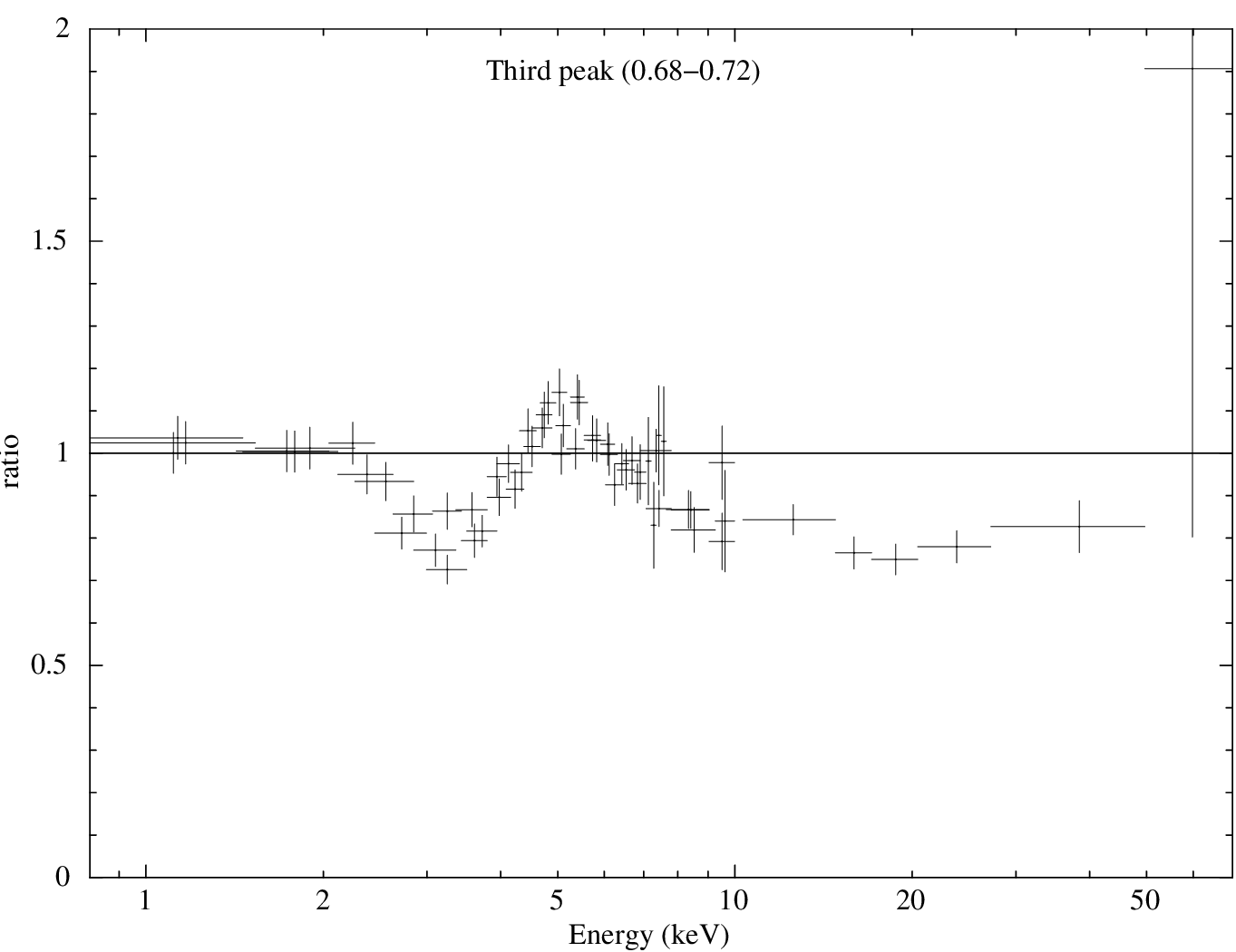}
\includegraphics[scale=0.35,angle=-90]{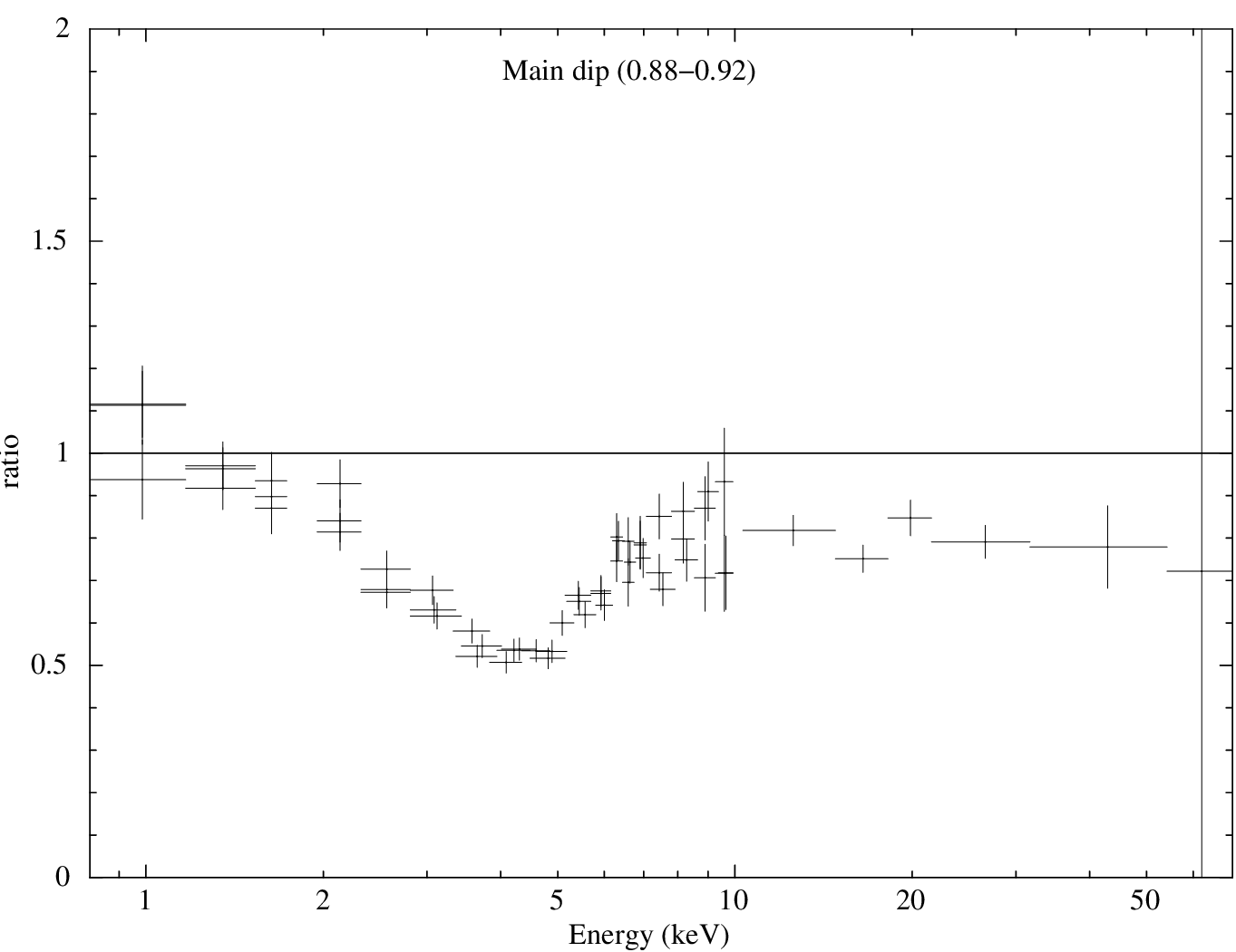}
\caption{Same as Figure 6 but now for Obs-2.}
\label{fig8}
\end{figure}

\begin{figure}
\includegraphics[height=12 cm, width=6 cm]{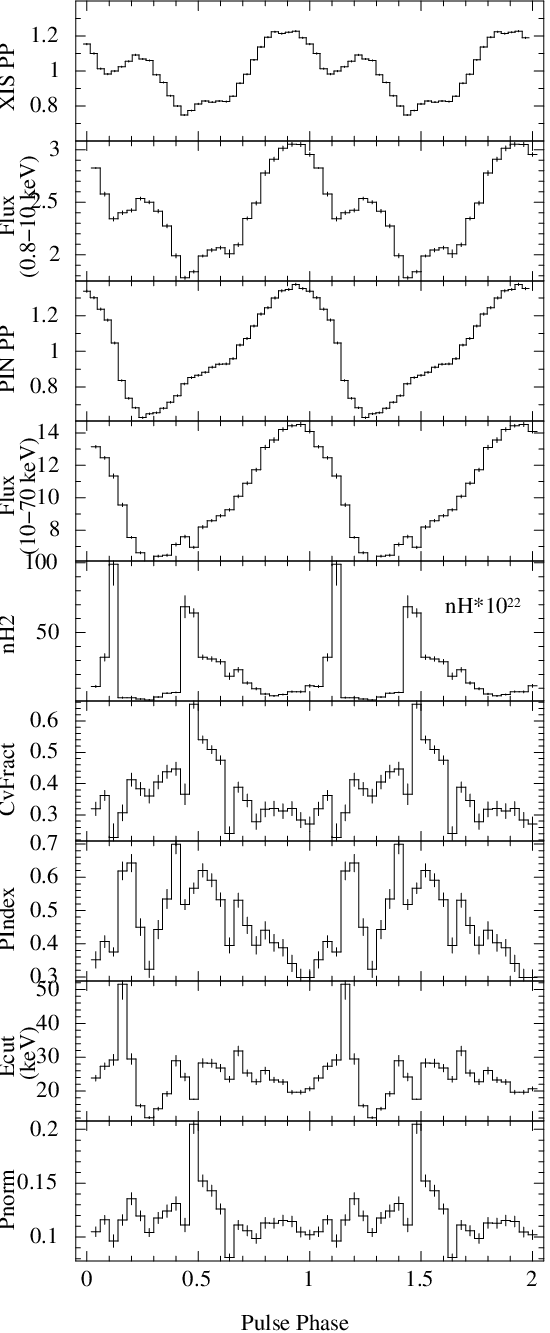}
\includegraphics[height=12 cm, width=6 cm]{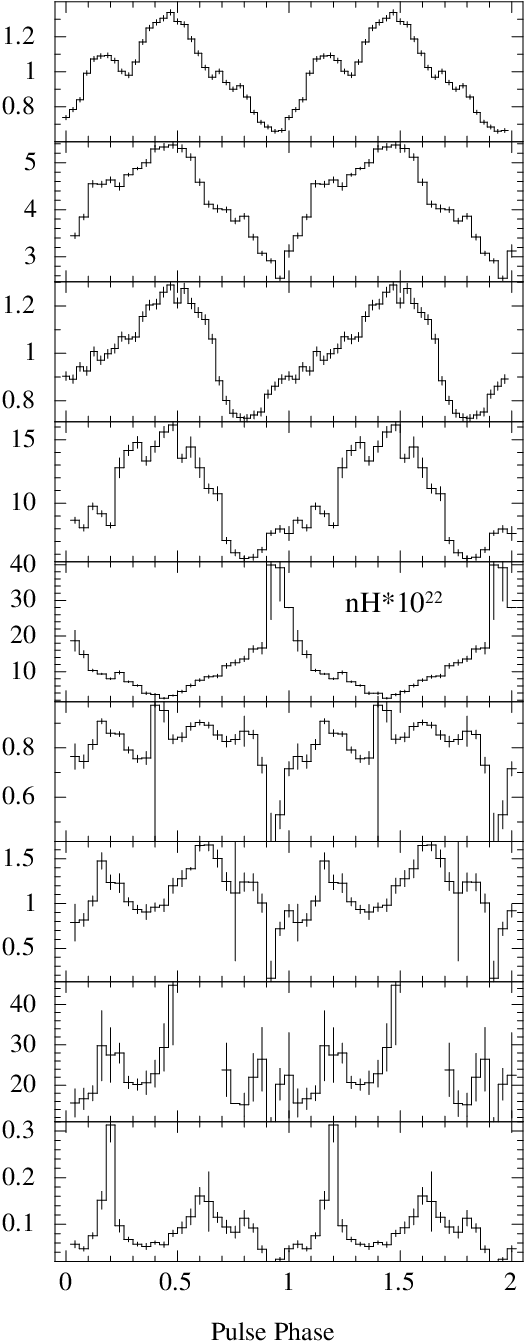}
\caption{Variation of the spectral parameters with phase along with the pulse profile in the 0.2--12 keV 
range (obtained from XIS data; top panels), source flux in the 0.8--10 keV range (second panels from top), 
pulse profile in the 10--70 keV range (obtained from PIN data; third panels from top) and source flux in 
the 10--70 keV range (fourth panels from top). The left panels show the variation of the spectral parameters
with pulse phase for Obs-1 and the right panels show the same for Obs-2. The units of flux and nH2 
(equivalent hydrogen column density) are in 10$^{-09}$ erg $cm^{-2} s^{-1}$ and $10^{22}$ atoms $cm^{-2}$,
respectively. Power-law normalization $Pnorm$ is in units of photons $keV^{-1} cm^{-2} s^{-1}$ at 
1 \rmfamily{keV}}.
\label{fig9}
\end{figure}

The strong energy dependence of the pulse profiles, as shown in Figures~2 \& 3 implies a dependence of the 
energy spectrum on the pulse phase. To investigate this, we performed pulse phase-resolved spectroscopy 
by dividing the data into 25 phase bins, applying phase filtering in the FTOOLS task XSELECT. Plotting the ratio 
of the spectra at some interesting phases, like the pulse peaks and dips to the phase average spectrum can 
clearly bring out the complexity of the spectral changes with pulse phase. In Figures~\ref{fig7} \&  \ref{fig8}, we plot the ratio of 
 few of the pulse phase-resolved spectra with the pulse phase averaged spectrum for both observations. 
For phase-resolved spectral analysis,For phase-resolved spectral analysis, we used same background spectra and response
matrices for the XIS and PIN data as was used in the case of phase
averaged spectroscopy.  We fitted the spectra in the same energy ranges and rebinned them by the same factor 
as in the phase averaged case for the XIS spectra. For the PIN spectra however, we rebinned Obs-1 by the same factor as the phase averaged spectra, but
 rebinned Obs-2 with a higher rebinning factor of 8 above 40 \rmfamily{keV} due to poor statistics. The value of $N_{H1}$ and the iron line width were fixed to the
corresponding phase averaged values (as given in Table~1). Due to limited statistics in the PIN energy range, 
we could not vary all the three parameters of the 'cyclabs' model simultaneously and hence fixed the cyclotron 
width to its phase averaged value. At the main dip of the pulse profiles, an additional peak at the lowest energy (less than 
2 \rmfamily{keV})
was apparent in the spectra of both observations, that indicate a soft excess and an additional blackbody 
component was required to fit the spectra for those phase ranges. For Obs-1, the phase bin corresponding to 0.40--0.44 required an 
additional blackbody component with a temperature $kT$=0.47 keV. Obs-2 on the other hand required a blackbody 
with a temperature $kT$=0.34 keV in the phase bin of 0.96--1.00. The radius 
$R_{bb}$ of the region from which this soft-excess is emitted can be calculated as \\ $\frac{L_{bb}}{4\pi R_{bb}^{2}}=\sigma T_{bb}^{4}$\\ Where $L_{bb}$ is the blackbody luminosity and $T_{bb}$ the corresponding temperature. The radius of 
the  blackbody emitting region determined from Obs-1 and Obs-2 are $\sim$ 6 \rmfamily{km} and 4 \rmfamily{km} respectively. The reduced $\chi^{2}$ for the phase-resolved spectroscopy were in the range of 1.03-1.46 
and 1.32-2.5 for 1024 degrees of freedom, for Obs-1 and Obs-2 respectively. 
The variation of the spectral parameters with pulse phase are shown in Figure~\ref{fig9}. The results obtained from the pulse phase resolved spectroscopy as
seen in this figure are summarized below: \\
 \textit{Obs-1 :}
There was an abrupt increase in the absorption column density $(N_{H2})$ in the narrow dip of the 
profile at phase $\sim$0.1 although the covering fraction ($Cv_{fract}$) remained small. In the main 
dip of the profile at pulse phase of $\sim$0.5, both the covering fraction and the $N_{H2}$ value increased. 
Spectral hardening was observed in both the peaks of the pulse profile.

\textit{Obs-2 :} 

  
The covering fraction remained nearly constant at a high value of $\sim$ 0.8 throughout the pulse phase 
indicating a more or less symmetric distribution of material around the neutron star. The value of
$N_{H2}$ increased dramatically at the main dip (phase $\sim$0.9), although the covering fraction 
decreased to $\sim$ 0.5. 
In the phase range $\sim$0.5--0.7 in Obs-2, no 
high-energy cutoff was required to fit the spectra. The reason for this is not very clear. 
\ \\
For both observations, the power-law continuum parameters like the power-law index (P$_{index}$), normalization (P$_{norm}$), and the cutoff energy (Ecut)
varied significantly with the pulse phase.

\begin{figure}
\includegraphics[scale=0.6,angle=-90]{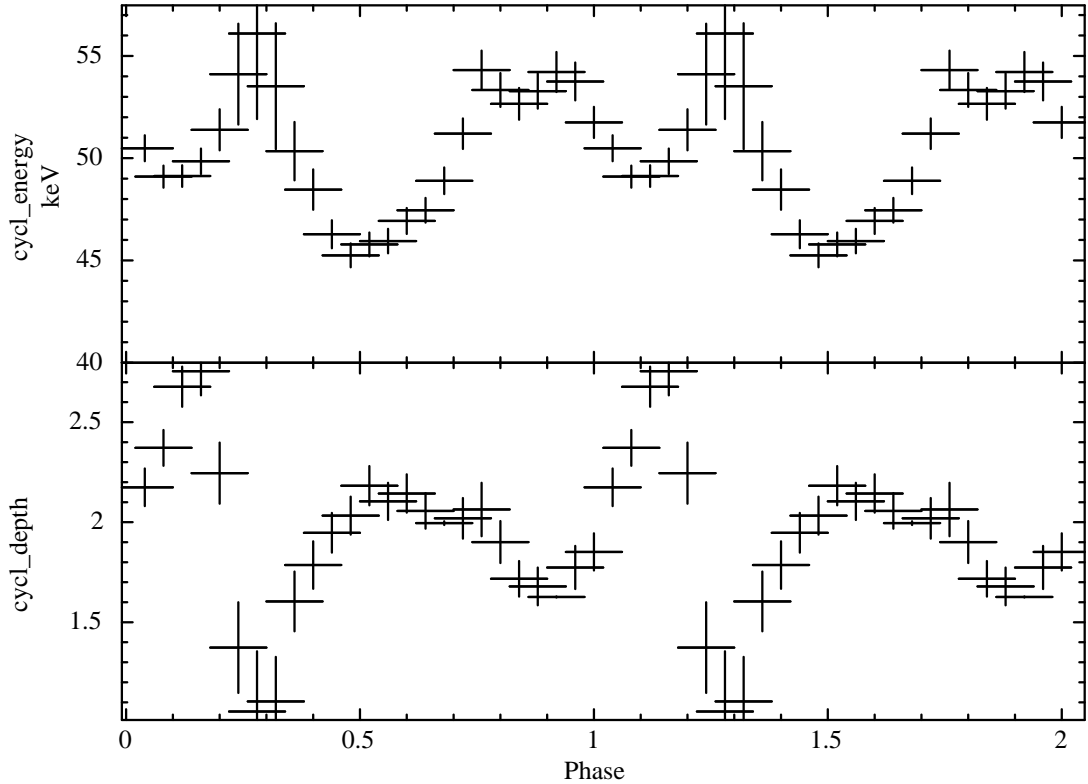}
\caption{Variation of the cyclotron energy and depth with the pulse phase for Obs-1. Only 8 of 
the 25 points are independent. cycl$\_$energy denotes the energy of the centroid of the CRSF feature \& cycl$\_$depth denotes the depth of the feature.}
\label{fig11}
\end{figure}

\textit{Phase resolved spectroscopy of the cyclotron line feature:}
\\
The high energy part (greater than 40 \rmfamily{keV}) of the data of Obs-1 had enough statistics to investigate the variation of cyclotron energy and depth
as a function of phase. Instead of varying all the three cyclotron parameters, we fixed
the cyclotron width to the phase averaged value.
 To perform phase-resolved spectroscopy of the cyclotron feature, 25 phase bins were used for the XIS data as in the previous cases. Due to limited statistics in the PIN data, pulse phase resolved PIN spectra with phases centered at the XIS phase bins were generated but at
thrice their widths, resulting in 25 overlapping phase bins out of which only 8 were independent. 
Figure~\ref{fig11} shows the variation of the cyclotron energy and depth with phase. There is 
an indication of anti-correlation or a phase shift between the energy and the depth of the CRSF feature. 
The cyclotron energy shows a variation of about 10 \rmfamily{keV} and has two peaks at the same phase of the pulse peaks in the XIS energy band. As is also evident from Figure~\ref{fig11}, The depth of the CRSF also shows a 
double peaked structure with large variation by a factor of 3 in a narrow phase interval of about 0.1.

\section{Discussion \& Conclusions}
We performed a detailed timing and broad-band pulse phase-resolved spectral analysis 
of the two $Suzaku$ observations of the HMXB transient pulsar 1A~1118-61 during its 2009 giant outburst in its peak (Obs-1) 
and declining phase (Obs-2).
Owing to the broad-band capability and 
high sensitivity of \emph{Suzaku}, we have been able to investigate the complex energy dependence of the pulse 
profiles in more detail and in narrow energy bins compared to the previous works. In 
addition, we have detected a soft-excess peak (in the 0.3--2 keV range) at the main dip of the pulse profiles 
for both observations. Broad-band spectral analysis of the \emph{RXTE} observations have been 
performed by \citet {doroshenko2011}, \citet {devasia2011} \& \citet {nespoli2011} who reported 
the presence of a CRSF at $\sim$55 keV. \citet {suchy2011} performed broad-band spectral analysis 
of the same \emph{Suzaku} observations using partial covering cutoff power law and compTT models and  
confirmed the presence of the CRSF feature at $\sim$55 keV, discussing also the possibility of the 
presence of a first harmonic at $\sim$110 keV. \citet{suchy2011} also included pulse phase-resolved analysis 
in three phase bins in the two peaks and in the minima of the pulse profiles where they reported
some variations in the continuum and the centroid of the cyclotron line with pulse phase. We  
carried out a more detailed phase-resolved spectral analysis of the \emph{Suzaku} observations in 
narrow phase bins using a partial covering cutoff power law model. Since the aim of the paper is to 
mainly investigate the variations of the spectral parameters with the pulse phase, we have not included 
GSO data in our analysis as the data quality is not suitable to perform phase-resolved analysis in 
narrow phase bins. We, therefore, cannot investigate the possible presence of a harmonic 
at $\sim$110 keV. Variations of the CRSF parameters over the pulse phase are brought out very clearly 
in this source, and both the energy and the depth of the CRSF shows some interesting patterns
of variation with pulse phase.
\subsection{Pulse Profiles: Energy Dependence \& Soft-Excess Peak}
Energy dependence of the pulse profiles have been found previously in other accretion powered X-ray pulsars such as 4U~0115+63 \citep {tsygankov2007} and Her~X-1 \citep {nagase1989}.
The energy dependence of the pulse profiles of the Be HMXB pulsar 1A~1118-61 during its 
2009 outburst shows a multi peaked structure below $\sim$ 10 \rmfamily{keV}, and a single asymmetric peak 
thereafter. These features have been reported previously using \emph{RXTE} observations by \citet {doroshenko2011}, \citet {devasia2011}, and 
\citet {nespoli2011}.  
Owing to the better energy resolution, and the broader energy coverage, especially below 3 keV, the 
$Suzaku$ data show a more detailed and clearer picture of the energy dependence of the pulse profiles. \citet{suchy2011} analyzed \emph{Suzaku} data 
using much wider energy bins than ours to probe the energy dependence.  
 The presence of dips in the pulse profiles 
can be explained due to the presence of an additional absorption component at that pulse phase which 
obscures the radiation \citep{galloway2001}. Recently similar energy dependence of the pulse profiles have also been found in other HMXB pulsars for 
example in GRO~J1008-57 \citep {naik2011}, and GX~304-1 \citep {devasia2011a}, where the presence of the dips in 
the pulse profiles could be explained due to the presence of additional 
absorption component at those phases.  \\
 We have detected a soft excess peak in the lowest energies (0.3--2 keV range) at the main dips of the pulse 
profiles for both observations. A soft excess emission is common among accretion powered pulsars that 
have a low line of sight absorption \citep {paul2002,hickox2004}. The presence of a soft excess in 
accreting X-ray pulsars can be explained as by reprocessing in the thick magnetosphere 
shell \citep {cray1982} or in the accretion column. The soft-excess could also be due to
the  reprocessing of the hard X-rays by the optically 
thick material in the inner edge of the accretion disk \citep {endo2000,ramsay2002}. In this work, 
fitting a blackbody component to the spectra of the phases corresponding to the soft excess peaks, we derived 
the radius of the soft excess emitting regions to be $\sim$ 6 km and $\sim$ 4 km for Obs-1 and Obs-2 respectively. The 
small radii in both the cases suggest that the soft excess is being produced  
in the accretion mound itself or as reprocessed emission from the accretion column. Its origin from 
the inner accretion disk is very unlikely since the size of the emission region is very small compared 
to the inner disc radius which is of a few hundred kms. Also, the soft excess produced by the reprocessed disc
blackbody component is usually seen over a broad phase range in other pulsars \citep{paul2002, naik2004} 
and not confined to a narrow phase range like in this source.
.
\subsection{Quasi Periodic Oscillations} 
The results related to the QPO feature found at $0.087\pm0.009$ Hz is almost similar to the one found by 
\citet{devasia2011} including its energy dependence. If the broad feature at $0.026\pm0.003$ Hz detected 
in Obs-2 is interpreted as being a QPO, assuming the generation of the QPO at the inner radius 
of the accretion disc, the decrease in the QPO frequency is consistent with the 
decrease in one order of magnitude of X-ray luminosity from Obs-1 to Obs-2 ($\nu_{qpo} \propto L_{x}^{3/7}$) . Using this argument, the inferred value of the magnetic field is also in 
agreement with the one found by the CRSF \citep{devasia2011}. The quality factor of the QPO feature 
in Obs-2 is 
however very low compared to the values found for the other accretion powered pulsars.

\subsection{Pulse Phase averaged Spectroscopy}
The strong and complex energy dependence of the pulse profiles bring out the requirement of a multicomponent model to fit 
the energy spectra of 1A~1118-61.
Some of the continuum models 
that have been used extensively to describe the spectra successfully  are the Negative \& Positive 
powerlaws with a common Exponential (NPEX) continuum model \citep {miharaa, makishima1999, terada2006, naik2008} 
and power-law with smooth Fermi-Dirac cutoff energy \citep{tanaka1986}. The partial covering absorption model 
\citep {endo2000, mukherjee2004} which consists of two power-law continua with a common photon index but 
different absorption components 
also describes the spectrum of these sources well. 
 We had initially tried to fit the data with a partial covering model with a high-energy cutoff, but 
the spectra in some phases had a very low value of cutoff energy ($\sim$ 0--3 \rmfamily{keV}) which is 
equivalent to a cutoff power-law model. The phase averaged data was thus well fitted with a partial covering cutoff power law model.
The spectrum was softer and steeper at the declining phase of 
the outburst similar to what was found in some other HMXB transients like V~0332+53 \citep {mowlavi2006} and 
A~0535+26 \citep {caballero2009}. A narrow Iron $\mathrm{K } \alpha$ line centered around 6.4 keV 
was detected in both observations which implied the presence of neutral or weakly ionized Iron. 
The possible origin of the iron line is the accretion disk or the circumstellar material, or the cool dense matter in the magnetosphere which reprocesses the emission.

The presence of the CRSF feature in the spectra had been reported previously by \citet {doroshenko2011}, 
\citet {nespoli2011}, $\&$ \citet {suchy2011}. From the spectral fitting with Lorentzian profiles, we found that 
the CRSF line centers at  $49 \pm 0.5$ and $52^{+4.9}_{-3.1} $ \rmfamily{keV} for Obs-1 and Obs-2 respectively. The 
best fitting centriods for Obs-1 and Obs-2 are consistent within confidence limits. The result is also consistent with a slight increase in the centroid 
energy with decreasing luminosity which is in agreement with the 
variation 
of the cyclotron line energy with that of the height of the accretion column. Higher the luminosity, 
lower in energy is the CRSF produced. This is because, higher luminosity implies higher accretion rate and a 
higher height for the accretion column. The CRSF feature is thus produced further away from the surface of the 
neutron star \citep {miharab, nakajima2006}. The CRSF feature is also broader in Obs-1 
which may be due to the superposition of CRSF features produced in scattering regions at different heights of 
the accretion column which is higher in this case than in Obs-2. 
\subsection{Pulse Phase resolved spectroscopy}
In addition to pulse phase averaged spectroscopy, we have performed pulse phase-resolved spectroscopy 
to understand the complex energy dependence of the pulse profiles. The column density of the 
material local to the neutron star ($N_{H2}$) 
of $\sim 10^{24} \quad atoms \quad cm^{-2}$ can explain the dips seen in the pulse profiles. The changes in the value of $N_{H2}$ and the covering fraction with pulse phase give us an idea about the 
properties of the plasma in the accretion 
stream. The plasma  may be a clumpy structure with different values of opacity and optical depth. We 
find that the accretion column in Obs-1 represents a more clumpy structure with the dip at phase $\sim$0.1 having a very high value of $N_{H2}$ ($\sim 10^{24} \quad 
atoms \quad cm^{-2}$) but a low covering value ($\sim$ 0.4). However, both values remain high in
 the main dip. Further, the accretion 
column in Obs-2, shows a more symmetric distribution 
of matter with a more or less constant value of covering fraction throughout the pulse phase.

The spectral hardening at the peaks of the pulse profile in Obs-1 may represent 
the passage of the magnetic axis through our line of sight, and a deep and a more direct view into the emission region 
along the magnetic axis \citep {pravdoa}. Similar regions of pulse hardening have been observed previously in other sources 
like Her~X-1 \citep {pravdob, pravdoc}, 4U~0115+63 \citep {johnston1978, rose1979}, \& GX 1+4 \citep {doty1981}.\\

Another feature that has been found is the variation of the CRSF parameters with pulse phase. Since the cross section and the depth of the
 cyclotron scattering feature is thought to depend significantly on the viewing angle of the accretion column, the CRSF properties
 may change with the rotation of the neutron star, hence with pulse phase. A large variation of $\sim$ 10 \rmfamily{keV} in the 
cyclotron energy and a factor of 3 variation in the depth could either imply the dipole geometry being viewed at different angles at different pulse 
phases, or a more complex underlying magnetic field structure. Fitting of this observed variation of the cyclotron parameters with 
phase with a detailed and suitable theoretical model can be used to infer the structure of the magnetic field around the neutron star 
and also put constraints on its other parameters like the inclination between the magnetic and rotation axis \rmfamily{etc.} This 
is however beyond the scope of the present work.

\section*{Acknowledgments}
We thank the reviewer for a very extensive review with many specific
suggestions that helped us to improve the paper.
This research has made use of data obtained through the High Energy Astrophysics Science Archive Research Center 
On line Service, provided by NASA/Goddard Space Flight Center.

\label{lastpage}

\end{document}